\documentclass[%
 aip,
 amsmath,amssymb,
 reprint,%
nofootinbib,
]{revtex4-1}

\usepackage{graphicx}
\usepackage{dcolumn}
\usepackage{bm}

\usepackage[utf8]{inputenc}
\usepackage[T1]{fontenc}
\usepackage{mathptmx}
\usepackage{etoolbox}
\usepackage{graphicx}
\usepackage{dcolumn}
\usepackage{bm}
\usepackage{hyperref}
\usepackage{physics}
\usepackage{siunitx}
\usepackage[dvipsnames]{xcolor}
\usepackage{empheq}
\usepackage{caption}
\usepackage{subcaption}
\usepackage{soul}

\makeatletter
\def\@email#1#2{%
 \endgroup
 \patchcmd{\titleblock@produce}
  {\frontmatter@RRAPformat}
  {\frontmatter@RRAPformat{\produce@RRAP{*#1\href{mailto:#2}{#2}}}\frontmatter@RRAPformat}
  {}{}
}%
\makeatother
\newcommand{\iu}{{i\mkern1mu}}

\renewcommand{\vec}{\boldsymbol}
\begin{document}

\preprint{AIP/123-QED}

\title{New Class of Landau Levels and Hall Phases in a 2D Electron Gas Subject to an Inhomogeneous Magnetic Field: An Analytic Solution
}
\author{Dominik Sidler}
  \email{dsidler@mpsd.mpg.de}
 \affiliation{Max Planck Institute for the Structure and Dynamics of Matter and Center for Free-Electron Laser Science, Luruper Chaussee 149, 22761 Hamburg, Germany}
\affiliation{The Hamburg Center for Ultrafast Imaging, Luruper Chaussee 149, 22761 Hamburg, Germany}

\author{Vasil Rokaj}
  \email{vasil.rokaj@cfa.harvard.edu}
    \affiliation{Max Planck Institute for the Structure and Dynamics of Matter and Center for Free-Electron Laser Science, Luruper Chaussee 149, 22761 Hamburg, Germany}
  \affiliation{ITAMP, Harvard-Smithsonian Center for Astrophysics, Cambridge, MA 02138, USA}

\author{Michael Ruggenthaler}
  \email{michael.ruggenthaler@mpsd.mpg.de}
  \affiliation{Max Planck Institute for the Structure and Dynamics of Matter and Center for Free-Electron Laser Science, Luruper Chaussee 149, 22761 Hamburg, Germany}
  \affiliation{The Hamburg Center for Ultrafast Imaging, Luruper Chaussee 149, 22761 Hamburg, Germany}

\author{Angel Rubio}
  \email{angel.rubio@mpsd.mpg.de}
  \affiliation{Max Planck Institute for the Structure and Dynamics of Matter and Center for Free-Electron Laser Science, Luruper Chaussee 149, 22761 Hamburg, Germany}
    \affiliation{The Hamburg Center for Ultrafast Imaging, Luruper Chaussee 149, 22761 Hamburg, Germany}
  \affiliation{Center for Computational Quantum Physics, Flatiron Institute, 162 5th Avenue, New York, NY 10010, USA}
  \affiliation{Nano-Bio Spectroscopy Group, University of the Basque Country (UPV/EHU), 20018 San Sebasti\'an, Spain}

\date{\today}

\begin{abstract}
An analytic closed form solution is derived for the bound states of a two dimensional electron gas subject to a static, inhomogeneous ($1/r$ in plane decaying) magnetic field, including the Zeeman interaction. The solution provides access to many-body properties of a two-dimensional, non-interacting, electron gas in the thermodynamic limit. Radially distorted Landau levels can be identified as well as magnetic field induced density and current oscillations close to the magnetic impurity. 
These radially localised oscillations depend strongly on the coupling of the spin to the magnetic field, which give rise to non-trivial spin currents. Moreover, the Zeeman interaction introduces a unique flat band, i.e. infinitely degenerate energy level in the ground state, assuming a spin $g_s$-factor of two. Surprisingly,  the charge and current densities can be computed analytically for this fully filled flat band in the thermodynamic limit.
Numerical calculations show that the total magnetic response of the electron gas remains diamagnetic (similar to Landau levels) independent of the Fermi energy. However, the contribution of certain, infinitely degenerate energy levels may become paramagnetic. Furthermore, numerical computations of the Hall conductivity reveal asymptotic properties of the electron gas, which are  driven by the anisotropy of the vector potential instead of the magnetic field, i.e. become independent of spin. Eventually, the distorted Landau levels give rise to negative and positive Hall conductivity phases, with sharp transitions at specific Fermi energies. Overall, our work merges ``impurity'' with Landau-level physics, which provides novel physical insights, not only locally, but also in the asymptotic limit. This paves the way for a large number of future theoretical as well as experimental investigations, e.g. to include electronic correlations and to investigate two-dimensional systems such as graphene or transition metal dichalcogenides under the influence of inhomogeneous magnetic fields.
\end{abstract}

\maketitle


\section{Introduction}

Lev Landau's analytic solution for the non-interacting electrons subject to a constant magnetic field, known as Landau levels, has served as a paradigmatic model system in condensed matter physics for almost a century.\cite{landau1930diamagnetismus, Landau} Its basic concepts are the foundation of numerous groundbreaking discoveries. To mention a few, the integer\cite{klitzing1980new} and fractional\cite{tsui1982two, Laughlingfractional} quantum Hall effect  are fundamentally related to the emergence of quantized Landau levels for a two dimensional electron gas in a homogeneous magnetic field.  Further, in the presence of a periodic potential, the Landau levels develop mini-gaps and for the energy spectrum a self-similar fractal pattern emerges, known as the Hofstadter butterfly,~\cite{hofstadter1976energy} which has become experimentally accessible via magnetotransprot measurements in Moir\'{e} materials.~\cite{DeanButterfly, BarrierButterfly, ForsytheButterfly, WangButterfly}
The study of Landau levels and topological edge states is still very actively pursued and is currently even considered in quantum optics and cavity quantum electrodynamics (QED), where ultra-strong coupling of the Landau levels to the quantum vacuum fluctuations  and the control of conduction properties have been achieved experimentally in a cavity,~\cite{ScalariScience, li2018, FaistCavityHall, paravacini2019,rubio2022new} with several theoretical studies and proposals accompanying these developments.~\cite{Hagenmuller2010cyclotron, CiutiHopping, rokaj2019, RokajButterfly2021,de2021light}

In parallel to the fundamental investigation of quantum systems exposed to magentic fields, the study of impurity models has a long lasting tradition in solid state physics. Some of the fundamental theoretical concepts go back to work of Friedel for charge impurity induced oscillations\cite{friedel1954electronic}, whereas Anderson localisation\cite{anderson1958absence} or Kondo effect\cite{kondo1964resistance} may emerge due to lattice induced magnetic impurities. Quantum impurity models are basic to nanoscience as representations of quantum dots and molecular conductors,\cite{hanson2007spins,gull2011continuous} or they have for example been used to understand the adsorption of atoms onto surfaces.\cite{brako1981slowly,langreth1991derivation,gull2011continuous} For quantitative predictions of atomic or molecular impurities, powerful computational methods are nowadays available (e.g. continuous-time Monte Carlo\cite{werner2006continuous, gull2011continuous}). 

In the following, we will introduce a fundamental theoretical model, which connects the world of magnetic impurities with the Landau setting in a non-perturbative way.
In more detail, we will derive a simple closed form solution for an electron subject to a radial symmetric $1/r$-decaying magnetic field including the spin-dependent Zeeman interaction. Our solution of the Pauli equation will serve as a fundamental ingredient to study spin-resolved local and asymptotic properties of a non-interacting 2D electron gas subject to a radially symmetric defect, which is induced by the externally applied magnetic field.   

The manuscript is structured as follows. In a first step, we derive the analytic boundstate solution for our inhomogeneous field setup. In a second step, single-electron properties are discussed with their implications for the consecutive many-body solution. Based on those considerations, local (magnetic field driven) properties (charge, current and magnetisation densities) of the electron gas are investigated analytically as well as numerically. Eventually, asymptotic (vector potential driven) Hall conductivities  can be infered for different type of electric field perturbations, based on locally converged numerical data. Finally, a brief outlook of the various implications of our exact solution is provided for different future research directions.

\section{Analytic Single Electron Solution \label{sec:anasol}}

As a starting point for our investigation of a non-interacting 2D electron gas subject to a perpendicular, radially symmetric, $1/r$ in plane decaying, static magnetic field $\vec{B}(\boldsymbol{r})$, we rely on the minimal coupling Hamiltonian operator in Coulomb gauge including the Zeeman interaction,
\begin{eqnarray}
\hat{H}&=&\sum_{j=1}^N\frac{1}{2m}\big[\hat{\vec{p}}_j-q \vec{A}(\boldsymbol{r}_j)\big]^2+\frac{g_s\mu_B}{2} \vec{\sigma}_j\cdot\vec{B}(\vec{r}_j).
\end{eqnarray}
The electron mass is indicated by $m$ with negative unit charge $q=-e$. We denote the usual canonical position operator of particle $j$ as $\vec{r}_j$ and the corresponding momentum operator as $\vec{p}_j$  and the anisotropic external vector potential is denoted by $\vec{A}(\boldsymbol{r})$. The Bohr magneton is indicated by $\mu_B=(e\hbar)/(2 m)$ and for the spin g-factor we assume the non-relativistic value $g_s=2$ throughout this work. The Pauli vector for electron $j$ is labeled by $\vec{\sigma}_j$.

In a next step, we define the external anisotropic vector potential within cylindrical coordinates, 
\begin{eqnarray}
\vec{A}(\boldsymbol{r}):=A_\phi \vec{e}_{\phi}
\end{eqnarray}
such that it assumes a constant value ($A_\phi =\rm const$) throughout space, with $\vec{e}_{\phi}=\frac{1}{r}(-y \vec{e}_x+x\vec{e}_y)$ indicating the unit vector along $\phi$-direction. The corresponding SI units are [$\rm Tm$] and it obeys the Coulomb gauge condition $\vec{\nabla}\cdot \vec{A}=0$. The corresponding inhomogeneous magnetic field is given by
\begin{eqnarray}
    \vec{B}(\boldsymbol{r})&=&\vec{\nabla}\wedge \vec{A}=\frac{1}{r}\frac{\partial(r A_\phi)}{\partial r}\vec{e}_z=\frac{A_\phi}{r}\vec{e}_z.\label{eq:B}
\end{eqnarray} 
Throughout the following derivation we assume $A_\phi<0$, which corresponds to an inhomogeneous magnetic field directed in negative $z$-direction.
The permeability of the free space is given by $\mu_0$.  
Notice, our inhomogeneous magnetic field corresponds to a radial external  current density of the form $J_{\mathrm{ext}}=\frac{\vec{\nabla}\wedge \vec{B}}{\mu_0}=
\frac{A_\phi}{\mu_0}\frac{1}{ r^2} \vec{e}_\phi$ if considered in free space. We will comment on other options later in Sec.~\ref{sec:intermediate}.

Having made these preliminary definitions, we can rewrite the electronic Hamiltonian operator in a more convenient form that eventually  provides access to its simple closed form solution.
The corresponding Hamiltonian of a non-interacting 2D electron gas, coupled to the classical $\vec{A}(\boldsymbol{r})$ and $\vec{B}(\boldsymbol{r})$ fields, is given by,
\begin{eqnarray}
\hat{H}&=&\sum_{j=1}^N \bigg[-\frac{\hbar^2}{2m}\vec{\nabla}^2_j+ \frac{A_\phi q\hbar}{m}\bigg(\iu  \frac{\partial}{r_j\partial\phi_j} -\frac{\sigma_{z,j}}{2r_j}\bigg) \bigg]+N\frac{q^2 A_\phi^2 }{2 m},\label{eq:clasham}\nonumber\\
\end{eqnarray}
in the $(r,\phi)$-plane.
Fortunately, the contribution of the diamagnetic term $ E_{A^2}:=\frac{q^2 A_\phi^2 }{2 m}$ remains constant for all $N$ electrons in radial coordinates, which reduces the complexity of our problem considerably. We would like to mention however, that for a quantized field the diamagnetic $\mathbf{A}^2$-term does not contribute just a constant energy per particle, but modifies drastically the spectrum and excitations of the electron-photon system.~\cite{RokajCavityGas}
In a next step, we introduce
\begin{eqnarray}
 \alpha:=\frac{A_\phi q \hbar}{m}>0,  
\end{eqnarray}
which allows a more compact notation for the following derivation. The resulting stationary Pauli equation for a single (!) electron can be written as
\begin{eqnarray}
\bigg[-\frac{\hbar^2}{2m}\Big(\frac{\partial^2}{\partial r^2}+\frac{\partial}{r \partial r}+\frac{\partial^2}{r^2\partial \phi^2}\Big)+ \iu \alpha \frac{\partial}{r\partial\phi}+\frac{\alpha\sigma_{z}}{2r}\bigg]\Psi&=&E \Psi, \label{eq:sg}\nonumber\\
\end{eqnarray}
where the constant $E_{A^2}$-term is neglected for the moment. Notice the close resemblance of Eq. (\ref{eq:sg}) to the two dimensional hydrogen atom. For this reason, similar solution strategies apply for our partial differential equation, as we will demonstrate subsequently.

The angular and spin problem can trivially be solved by separation of variables as $\Psi(r,\phi,s)=R(r)\Phi(\phi)\chi(s)$, with spin function $\chi$ and $\Phi=e^{\iu l \phi}$ with $l\in Z$, $s=\pm\frac{1}{2}$, since $[\hat{H},\frac{\partial}{\partial\phi_j}]=0$. This leaves us with the radial problem
\begin{eqnarray}
\hat{H}_{l,s}R&:=&
\bigg[-\frac{\hbar^2}{2m}\big(\frac{\partial^2}{\partial r^2}+\frac{\partial}{r \partial r}-\frac{l^2}{r^2}\big)-\alpha \frac{l+s}{r}\bigg]R=ER.\label{eq:radial}
\end{eqnarray}
Before continuing the solution of our radial problem, we need to distinguish two formally different cases: The interaction with the $B$-field gives rise to a Coulomb potential like $1/r$-term which is attractive if $l+s>0$ or repulsive if $l+s<0$ for a fixed $\alpha>0$. Notice that the third case $l+s=0$ cannot occur for spin-half particles, due to the  Zeeman interaction. 

\textbf{Bound states for $\boldsymbol{l+s> 0}$:} Let us now focus on the attractive eigenvalue problem given in Eq. (\ref{eq:radial}) with $l+s>0$.
Notice that from the positivity of the Laplacian (kinetic energy) operator 
$\hat{T}=\hat{H}_{l}(\alpha=0)=\frac{1}{2}(\hat{H}_{l+s>0}+\hat{H}_{l+s<0})$, we find the following relation  $\langle\hat{H}_{l+s>0}\rangle\leq\langle\hat{T}\rangle\leq\langle\hat{H}_{l+s<0}\rangle$. 
%
%
To solve the attractive eigenvalue problem  we apply the method of Frobenius and match orders of a series expansion. 
Therefore, we define 
\begin{eqnarray}
    \rho&:=&\sqrt{\frac{8 m|E|}{\hbar^2}}r,\label{eq:rhodef}\\
    \lambda_{l,s}&:=&\alpha (l+s) \sqrt{\frac{m}{2\hbar^2|E|}}>0,
\end{eqnarray}
for which our radial problem assumes a convenient form, \cite{boyle,schlicht}
\begin{eqnarray}
\bigg[\frac{\partial^2}{\partial \rho^2}+\frac{\partial}{\rho \partial \rho}-\frac{l^2}{\rho^2}+  \frac{\lambda_{l,s}}{\rho}-\frac{1}{4}\bigg]R(\rho)=0.\label{eq:radial_u}
\end{eqnarray}
To reach a simple closed form solution, we introduce the Ansatz $R(\rho)=e^{-\rho/2} f(\rho)$ in agreement with the literature~\cite{boyle}
\begin{eqnarray}
\bigg[\frac{\partial^2}{\partial \rho^2}-\frac{\partial}{\partial \rho}+\frac{\partial}{\rho \partial \rho}-\frac{l^2}{\rho^2}+  \Big(\lambda_{l,s}-\frac{1}{2}\Big)\frac{1}{\rho}\bigg]f(\rho)=0.\label{eq:radial_f}
\end{eqnarray}
The Ansatz is motivated, since for large $\rho$ our system approaches $\Big[\frac{\partial^2}{\partial \rho^2}-\frac{1}{4}\bigg]v(\rho)$, which has the normalisable solution $v(\rho)=e^{-\rho/2}$.
If we apply the series representation $f(\rho)=\sum_{i=0}^\infty c_i \rho^i$
and match the different orders in $\rho$, we find after an index shift $i\mapsto i+1$ with $c_{-1}=0$: 
\begin{eqnarray}
\sum_{i=-1}^\infty c_{i+1} i(i+1)\rho^{i-1}-c_i i\rho^{i-1}+c_{i+1} (i+1) \rho^{i-1} &&\\
- \frac{1}{2} c_i \rho^{i-1}-l^2 c_{i+1} \rho^{i-1}+\lambda_{l,s} c_i \rho^{i-1}&&=0\nonumber \label{eq:shiftedseries}
\end{eqnarray}
This gives  rise to the indicial equation:
\begin{eqnarray}
c_{i+1}[(i+1)^2-l^2]=c_i[i+\frac{1}{2}-\lambda_{l,s}].
\end{eqnarray}
It implies the "series switches on" for $c_{i+1}$ when $(i+1)^2=l^2$, i.e. $i+1=l$, and it can terminate only if $i+\frac{1}{2}=\lambda$. Otherwise one would converge to a non-normalizable solution since $c_{i+1}\rightarrow \frac{c_i}{i}$ for large $i$ and $f\rightarrow \sum_{i=0}^\infty\frac{\rho^i}{i !}$. Now, introducing quantum number $n:=i=\lambda-\frac{1}{2}$ leads to a simple closed form solution for the energy eigenvalues 
\begin{eqnarray}
E_{n,l,s}=-\frac{q^2 A_\phi^2
}{2m}\bigg[\frac{2(l+s)}{2n+1}\bigg]^2,\ n\geq l,\ l+s>0.\label{eq:Ebound}
\end{eqnarray}



Finally, reintroducing the initially neglected diamagnetic energy shift $E_{A^2}$ leads to the total  one-electron energy within an inhomogeneous, $1/r$-decaying magnetic field:
\begin{eqnarray}
\boxed{E^{\mathrm{tot}}_{n,l,s}=\frac{q^2 A_\phi^2
}{2m}\bigg(1-\bigg[\frac{ 2(l+s)}{2n+1}\bigg]^2\bigg),\ n\geq l,\ l+s>0.} \label{eq:Etotbound}
\end{eqnarray}
Notice that likewise, hydrogen related, quantization rules arise for two dimensional magnetic quantum dots.\cite{downing2016magnetic,downing2016massless} However, in the following we can construct the respective explicit closed form eigenfunctions, which will provide fundamental physical insights not only analytically but also numerically.

\textit{Remark:} The exact solution of the non-attractive eigenvalue problem $(l+s\leq 0)$ will remain unknown, since the Frobenius method does not terminate anymore under these circumstances.

\textbf{Eigenfunctions:}
After having identified the energy eigenvalues for $n\geq l, l+s>0$, we can next find the corresponding eigenfunctions by expressing $f(\rho) = \rho^{l} L(\rho)$.\cite{schlicht} This turns Eq.~\eqref{eq:radial_f} into
\begin{eqnarray}
\rho \frac{d ^2 L}{d\rho^2}+(\nu+1-\rho)\frac{dL}{d\rho}+w L&=&0,\ w,\nu\in \mathbb{N}_0,\label{eq:lagid}
\end{eqnarray}
which can be solved by the associated Laguerre polynomials $L_w^\nu$ of degree $w$ and parameter $\nu$.~\cite{arfken1999mathematical} The associated Laguerre polynomials are given by Rodrigues' formula,\cite{arfken1999mathematical}
\begin{eqnarray}
L_w^\nu(\rho)=\frac{\rho^{-\nu}e^\rho}{w!}\frac{d^w}{dx^w}(e^{-\rho}\rho^{w+\nu}).
\end{eqnarray}
It is straightforward to show that our radial Eq.~(\ref{eq:radial_u})  transforms into Eq. (\ref{eq:lagid}) for  the discovered energy eigenvalue $E_{n,l,s}$ with
\begin{eqnarray}
R_{n,l,s}=e^{-\frac{\rho}{2}}\rho^l L_{n-l}^{2l}(\rho).\label{eq:eigu}
\end{eqnarray}
Consequently, the orthonormal eigenfunctions of our full problem can be written as 
\begin{eqnarray}
\boxed{\Psi_{n,l,s}=\frac{1}{\sqrt{N_{n,l}}}e^{il\phi}e^{-\frac{\rho}{2}}\rho^l L_{n-l}^{2l}(\rho)\chi(s),}\label{eq:eigf}
\end{eqnarray}
which is identical to the 2D Hydrogen atom solution, except for a different energy scaling in radial coordinates, 
\begin{eqnarray}
\rho(r)=\frac{2 q A_\phi}{\hbar}\frac{2(l+s)}{2n+1}r,\label{eq:scaling}
\end{eqnarray}
which was introduced in Eq. (\ref{eq:rhodef}), 
The corresponding normalisation is explicitly calculated as,\cite{arfken1999mathematical}
\begin{eqnarray}
N_{n,l}&=&\int_0^{2\pi}\int_0^\infty e^{-\rho} \rho^{2l} \big(L_{n-l}^{2l})^2\rho d\rho d\phi\nonumber\\
&=&2\pi \frac{(n+l)!}{(n-l)!}(2n+1).
\end{eqnarray}
The orthogonality of the eigenfunctions  for different eigenvalues is easy to demonstrate. The orthogonality of degenerate eigenstates is a little more involved, yet will be shown subsequently. 
Notice also that the spin quantum number $s$ enters the energy dependent scaling factor $\rho$, which introduces a radial spin-dependence of the wave-function.  Different normalisation constants arising for $r$-dependent eigenfunctions are given in App.~\ref{app:ef}.

\textbf{Boundary conditions and uniqueness of eigenstates:}

Finally we comment on the uniqueness of the eigenfunctions and its boundary conditions. Since the radial equation is a second-order differential equation it allows for, in general, two linearly independent solutions. A unique solution is then usually either fixed by choosing appropriate boundary conditions or by normalizability. Since the different forms of the radial equations are of Sturm-Liouville type, there are very general results available that clarify which conditions are appropriate for a self-adjoint Hamiltonian~\cite{zettl2010sturm, bailey2001algorithm}. At $\rho \rightarrow \infty$ no boundary condition can be chosen and normalizability singles out the unique asymptotic form of the two linearly independent solutions $\nu_{\pm}(\rho) = e^{\pm \rho/2}$. At the lower endpoint $\rho = 0$ normalizability singles out the unique form if $l \neq 0$ in analogy to the three-dimensional hydrogen case.~\cite{teschl2014mathematical} This becomes apparent from Eq.~\eqref{eq:lagid}, for which a Sturm-Liouville classification of the different endpoints based on the value of $l$ exists.~\cite{bailey2001algorithm,everitt2005catalogue} Again, in analogy to the usual hydrogen case, for $l=0$ different boundary conditions at $\rho=0$ can be chosen, and we have selected the usual Friedrich's boundary conditions~\cite{bailey2001algorithm,everitt2005catalogue}.

\textbf{Analytic solutions for $g_s=0$:}

Notice that our analytic solution also applies for $g_s=0$, which is equivalent to no Stern-Gerlach term (neglected Zeeman interaction), i.e. the normal Schr\"odinger equation is solved instead of the Pauli equation.
For this reduced Hamiltonian, we find that the expressions for the eigenvalues and eigenstates given in Eqs. (\ref{eq:Etotbound}) and (\ref{eq:eigf}) still apply, when setting $s=0$ and $n\geq l>0$. However, despite this minor modifications, emerging from the interaction of the spin-$1/2$ particles with the magnetic field, local properties  can strongly deviate between $g_s=2$ and $g_s=0$, as will be shown in Sec. \ref{sec:egas}.


\section{Single Electron Properties and Distorted Landau Levels}

\begin{figure*}
     \centering
     \begin{subfigure}[b]{0.8\textwidth}
         \caption{}
         \centering
         \includegraphics[width=\textwidth]{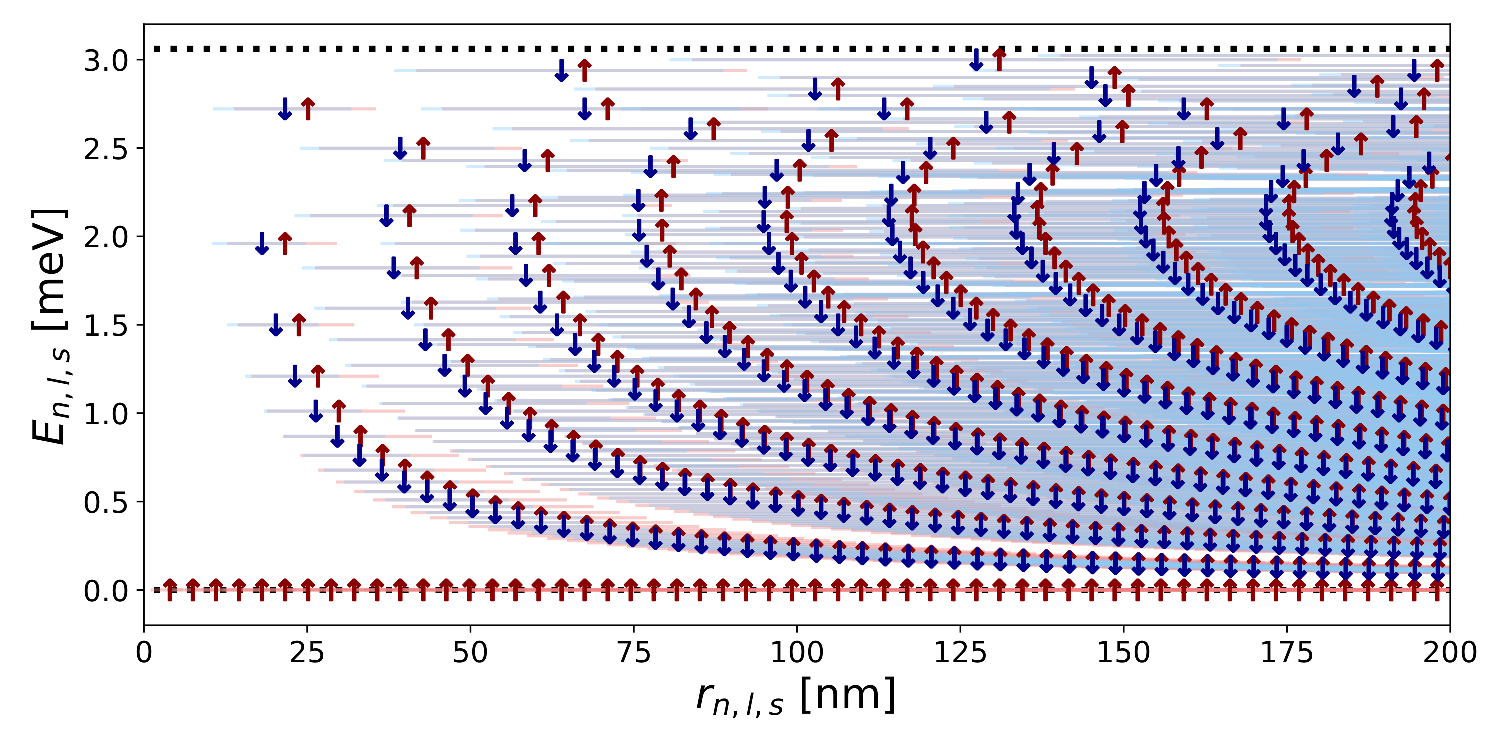}
         \label{fig:denEF0}
     \end{subfigure}
     \hfill
     \begin{subfigure}[b]{0.8\textwidth}
         \centering
         \caption{}
         \includegraphics[width=\textwidth]{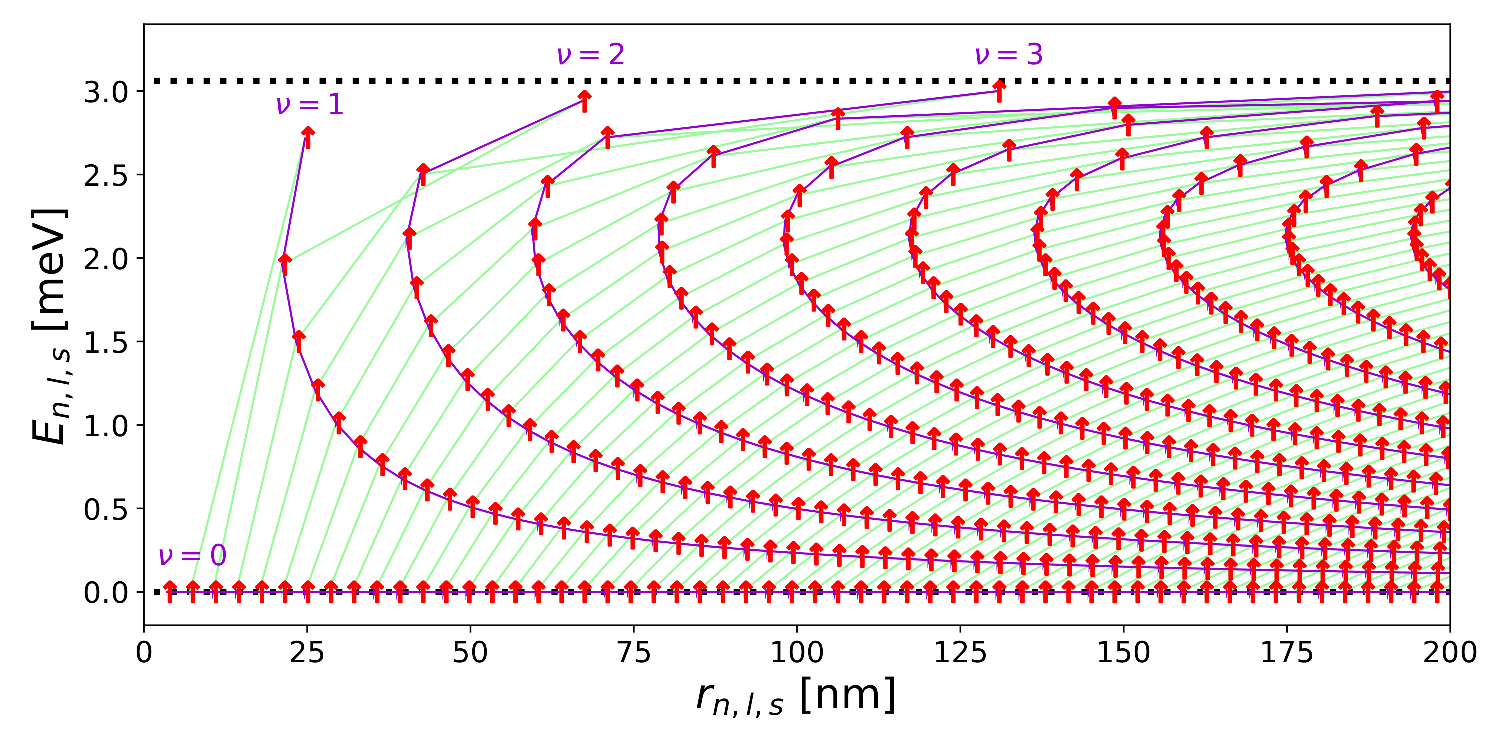}
         \label{fig:transitions}
     \end{subfigure}
        \caption{Radially resolved $\langle r\rangle_{n,l,s}$ energy eigenvalues $E_{n,l,s}$ of the electrons for a $A_\phi/r$-decaying magnetic field with $A_\phi=-0.186$ $\mu$Tm. 
        The black doted horizontal lines indicate the boundaries of allowed energy eigenvalues for our solution. Notice, that higher-lying energies exist for $l+s\leq0$, but the solutions remain unknown.
        In (a)  B-field aligned spin configuration are marked with $\color{blue}\boldsymbol{\downarrow}$ and anti-aligned electrons with $\color{red}\boldsymbol{\uparrow}$. Small horizontal lines correspond to the radial uncertainties $\Delta r=\sqrt{\langle r^2\rangle-\langle r\rangle^2}$. Overall, it is clearly shown that the presence of the inhomogeneous magnetic field increases the radial electron localisation close to the origin in combination with a coarse graining of the discrete energy levels.  Further away from the origin the denseness of the energy spectrum becomes apparent. The lowest lying ($E=0$), infinitely degenerate, electronic states can only emerge if the Zeeman interaction between the magnetic field and the spin is considered. In (b), the dipole allowed transitions are indicated by green lines for $\Delta n=0$ and purple lines for $\Delta n=1$. For illustrative purpose, we restrict our visualisation to the spin-$\color{red}\boldsymbol{\uparrow}$ states (equivalent results hold for the spin-$\color{blue}\boldsymbol{\downarrow}$ states). The first few distorted Landau levels (purple) are labeled by $\nu\geq0$ at their respective highest energies. Notice the strong curvature for $\nu\geq 1$. Similarly,  distorted Landau levels could also be identified for spin-down electrons starting at $\nu=1$.}
\end{figure*}

In a next step, we explore single electron properties based on our analytic solutions  in a $1/r$-decaying  magnetic field.
From the quantized energy given in Eqs.~(\ref{eq:Etotbound}) one can immediately derive fundamental spectral properties (see Fig.~\ref{fig:denEF0}):
\begin{enumerate}
\item \textbf{Bounded Energy Domain:} For the allowed quantum numbers it is straightforward to see  that  the constant diamagnetic energy shift $E_{A}$ compensates exactly for the attractive potential term entering Eq. (\ref{eq:radial}). Hence, we find  
\begin{eqnarray}
0
\leq E^{\mathrm{tot}}_{n,l,s}\leq E_{A^2}
.\label{eq:Ebound_domaintot}
\end{eqnarray}
This implies that the diamagnetic energy $E_{A^2}=\lim_{n\rightarrow\infty}E^{\mathrm{tot}}_{n,l,s}$ determines an $A_\phi$-dependent upper bound in our gauge choice, beyond which the unknown (!) solutions of the non-attractive eigenvalue problem commence. 
\item \textbf{Spin-half Degeneracy:} 
Each energy level is spin degenerate, i.e. $E^{\mathrm{tot}}_{n,l,1/2}=E^{\mathrm{tot}}_{n,l+1,-1/2}$, except for the ground-state energy $E^{\mathrm{tot}}_{n,n,1/2}=0$, which solely consists of spin up electrons, assuming a magnetic field direction  along the negative $z$-axis.
\item \textbf{Dense Energy Spectrum:} It is straightforward to show the dense nature of the energy spectrum given in Eq.~(\ref{eq:Etotbound}). Indeed every element in the interval given by Eq.~\eqref{eq:Ebound_domaintot} is a limit point, i.e. we have eigenvalues arbitrarily close.
\item \textbf{Infinite Energy Degeneracy:}
Interestingly,  in addition to the dense nature of the energy spectrum, one can also show that  each energy level is infinitely degenerate (as Landau levels are), since no restrictions apply to the  radial space for our solution, i.e. $\rho\in[0,\infty)$. The degeneracy becomes immediately evident for the lowest energy $E^{\mathrm{tot}}_{n,n,1/2}=0$. However,  it is a general property of each energy eigenvalue as one sees by setting $E^{\mathrm{tot}}_{n,l,s}\overset{!}{=}\rm{const}$, whose representative solution for the spin up case ($s=1/2$) is obtained from:   
\begin{eqnarray}
\frac{2l+1}{2n+1}=\frac{\mathrm{odd}}{\mathrm{odd}}=\mathrm{const},
\end{eqnarray}
with,
\begin{eqnarray}
l &=& \frac{(2l_0+1)k-1}{2}\label{eq:ldeg}\\ 
n&=&\frac{(2n_0+1)k-1}{2} \label{eq:deg_n}
\end{eqnarray}
and $\ k \in \{2D+1,\ D\in \mathbb{N}_0\}$ (similar solution applies for spin down $s=-1/2$ with $l\mapsto l+1$). The introduced subscript $0$ indicates the smallest allowed quantum numbers for a fixed energy eigenvalue. The relations to generate degenerate eigenvalues given in Eqs.~(\ref{eq:ldeg}) and (\ref{eq:deg_n}) have important consequences for the eigenfunctions defined in Eq.~(\ref{eq:eigf}). Indeed, they ensure that every degenerate eigenvalue (with identical spin) possesses a unique angular quantum number $l$. This automatically imposes orthogonality on the corresponding eigenstates via angular or spin selection rules. 
The infinite degeneracy of the energy spectrum is a very important property which connects directly our solution to the well-known Landau levels. Infinite degeneracy shows up also for the Landau levels, as the energy spectrum is independent of the momentum quantum number~\cite{landau1930diamagnetismus, Landau}. This fact is directly connected to the quantization of the Hall conductance~\cite{klitzing1980new}. In our inhomogeneous case however, the infinite degeneracy is much more intricate, than for the Landau levels, as it depends on the interplay between two quantum numbers, $n$ and $l$. 
%
\item\textbf{Exponential Localisation:} The radial exponentially suppressed  localisation of the eigenstates given in Eq. (\ref{eq:eigf}) has an interesting interpretation. First, by definition they correspond to a boundstate solution of our single electron Hamiltonian operator. Second, they can be considered as an ``intermediate'' regime $\sim \exp(-r)$ between the delocalised free electron gas solution $\sim 1$ and the more localised Gaussian eigenstates $\sim \exp(-r^2)$ of the Landau solution in homogeneous magnetic fields. Consequently, we anticipate that the influence of electron-electron interaction should overall be less severe for our inhomogeneous magnetic setup in comparison with the more localised Landau case and the subsequently developed non-interacting many-body solution should indeed represent a physically reasonable model. For our setting, the radial localisation of the eigenstates is visualised by means of radial uncertainties $\Delta r$ in Fig.~\ref{fig:denEF0}. It reveals that  the uncertainty increases considerably, particularly for larger energies and radii, and can spread over dozens of nm, i.e. can be considered strongly delocalised with respect to typical molecular scales.
\end{enumerate}



Having access to simple closed form solutions of the eigenfuctions in Eq.~(\ref{eq:eigf}) allows the determination of rigorous dipole selection rules for angular- and spin-quantum number. By evaluating $\bra{n,l,s}q\vec{r}\ket{n^\prime,l^\prime,s^\prime}$ one finds,
\begin{eqnarray}
\Delta l=\pm 1,\ \Delta s=0.\label{eq:diprulerig}
\end{eqnarray}
for dipole allowed transitions. Interestingly, numerical calculations of the radial integrals (see App. \ref{app:ef}) show that approximately even more stringent dipole selection rules apply,
\begin{eqnarray}
\Delta n\in \{0,1\} ,\ \Delta l=\pm 1,\ \Delta s=0.\label{eq:diprule}
\end{eqnarray}
which leads in for the allowed quantum numbers $n\geq l>0,\ l+s>0$ to effectively only two relevant transition channels for single electron states $n^\prime,l^\prime,s^\prime$, as visualised in Fig.~\ref{fig:transitions} for $s=1/2$. 
Indeed, the dipole allowed transition pattern visualised in Fig. \ref{fig:transitions} suggests the definition of \textbf{distorted Landau levels $\nu$} in our setting: 
\begin{eqnarray}
    \nu:=n-l,\ 
    \begin{cases}
\nu\in \mathbb{N}_0 &\text{if $s=1/2$}\\
\nu\in \mathbb{N} &\text{if $s=-1/2$}
\end{cases}
\end{eqnarray}
which are illustrated by the purple lines for $s=1/2$. The corresponding inter-level transitions (green) obey $\Delta\nu=\pm 1$, whereas the allowed intra-level transitions (purple) obey $\Delta \nu=0$. 
Notice that the lowest lying Level $\nu=0$ remains flat, as it is the case for Landau levels, provided that the Zeeman interaction is considered with $g_s=2$.
For $\nu>0$, the levels become strongly distorted with respect to radius $\langle r\rangle$. For relatively low energies (and large radii), they remain relatively flat, but having a negative curvature $dE/d\langle r\rangle\lessapprox0$, which becomes increasingly negative for intermediate energies, eventually culminating in a flat positive curvature for $E\lessapprox E_{A^2}$. The emergence of such strongly distorted Landau levels has important consequences for the Hall conductivity as will be seen in Sec. \ref{sec:hallcond}. Notice that lifted degeneracies of the Landau levels have previously been observed in special Landau settings with locally constant (but inhomogeneous) magnetic fields\cite{kim1996landau} or under the influence of additional electric fields.\cite{edery2019new}


\section{2D Electron Gas in Inhomogeneous Magnetic Field\label{sec:egas}}

After having discussed the single electron solution for our inhomogeneous magnetic field setting, we
continue by investigating fundamental many-body properties for non-interacting electrons. In other words, we leave the ``atomistic'' single electron perspective   and focus on Fermi energy $E_F$ dependent ``solid state'' characteristics instead. Surprisingly, we will find that the many-body problem cannot only be evaluated numerically, but there is even a simple closed form solution accessible in the thermodynamic limit $(N\rightarrow \infty)$ at $E_F=0^+$. This allows unique physical insights complementary to numerical calculations. Overall, we will focus on spatially sensitive properties (e.g. local densities), which are strongly affected by the $1/r$-decaying $B$-field at the origin, as well as asymptotic observables (e.g. Hall conductivity) that are dominated by the influence of the constant vector potential $A_\phi$ instead. Specifically the spatially-dependent properties (see also Figs~\ref{fig:denEF0} and \ref{fig:transitions}) will highlight a strong departure from the usual condensed-matter perspective and show how extended systems are connected to the more local atomic and molecular physics.

\subsection{Analytic Solution for the Distorted Landau Level at $\nu=0$}
We continue with the derivation of a simple closed form solution for the charge, current and magnetization densities of the fully filled lowest level $\nu=0$ ($E_F=0^+$).
For this purpose, we introduce the charge density,
\begin{eqnarray}
 \hat{n}(\vec{r})&=&q\sum_{i=1}^N \delta(\vec{r}-\vec{r}_i)\label{eq:chargensdef}
\end{eqnarray}
as well as the physical charge current density  in the Coulomb gauge,\cite{schafer2021making}
\begin{eqnarray}
 \vec{j}(\vec{r})&=& \vec{j}^{\rm orb}(\vec{r})+\vec{j}^{\rm s}(\vec{r})\label{eq:curent_tot}  
\end{eqnarray}
which is decomposed into orbital  $\vec{j}^{\rm orb}=\vec{j}^{\rm para}+\vec{j}^{\rm dia}$ current contributions, arising from the paramagnetic $\vec{j}^{\rm para}$ and diamagnetic $\vec{j}^{\rm dia}$ terms, and the spin-dependent magnetization current density $\vec{j}^{\rm s}(\vec{r})$ due to the Stern-Gerlach term. 
The different charge current density observables can explicitly defined as,
\begin{eqnarray}
\vec{j}^{\mathrm{para}}(\vec{r})&:=& \frac{\hbar q}{2mi}\sum_{i=1}^N \big(\delta(\vec{r}-\vec{r}_i) \overrightarrow{\nabla}_i- \overleftarrow{\nabla}_i \delta(\vec{r}-\vec{r}_i)\big)\label{eq:currdenspdef}\\
\vec{j}^{\mathrm{dia}}(\vec{r})&:=&-\frac{ q^2}{mc}\sum_{i=1}^N \delta(\vec{r}-\vec{r}_i)\vec{A}(\vec{r})\label{eq:currdensddef}\\
\vec{j}^{\mathrm{s}}(\vec{r})&:=&\frac{\hbar q}{2m}\sum_{i=1}^N\overrightarrow{\nabla} \times \vec{\sigma}_i\delta(\vec{r}-\vec{r}_i)
=\vec{\nabla}\times \vec{m}^s(\vec{r}).\label{eq:currdenssdef}
\end{eqnarray}

In the last step, the divergence free magnetization current density was expressed as the curl of the magnetization density $m^s$. It can be written in a particularly simple form for the non-interacting electrons of our system, 
\begin{eqnarray}
\vec{m}^s(\vec{r})&=&\frac{\hbar q }{m} \sum_{i=1}^N s_i \delta(\vec{r}-\vec{r}_i)\vec{e}_z.\label{eq:magn}
\end{eqnarray}
Notice that the origin of the magnetization current $\hat{\vec{j}}^{\mathrm{s}}$ is purely quantum mechanical, since it is spin-dependent, whereas orbital currents can also emerge in a classical setting. Furthermore, the magnetization current can only play a significant role for inhomogeneous spin magnetizations $m_z(\vec{r})$, which are not present in the ubiquitous Landau setting. 

\begin{figure}
     \centering
     \begin{subfigure}[b]{0.45\textwidth}
         \centering
         \includegraphics[width=\textwidth]{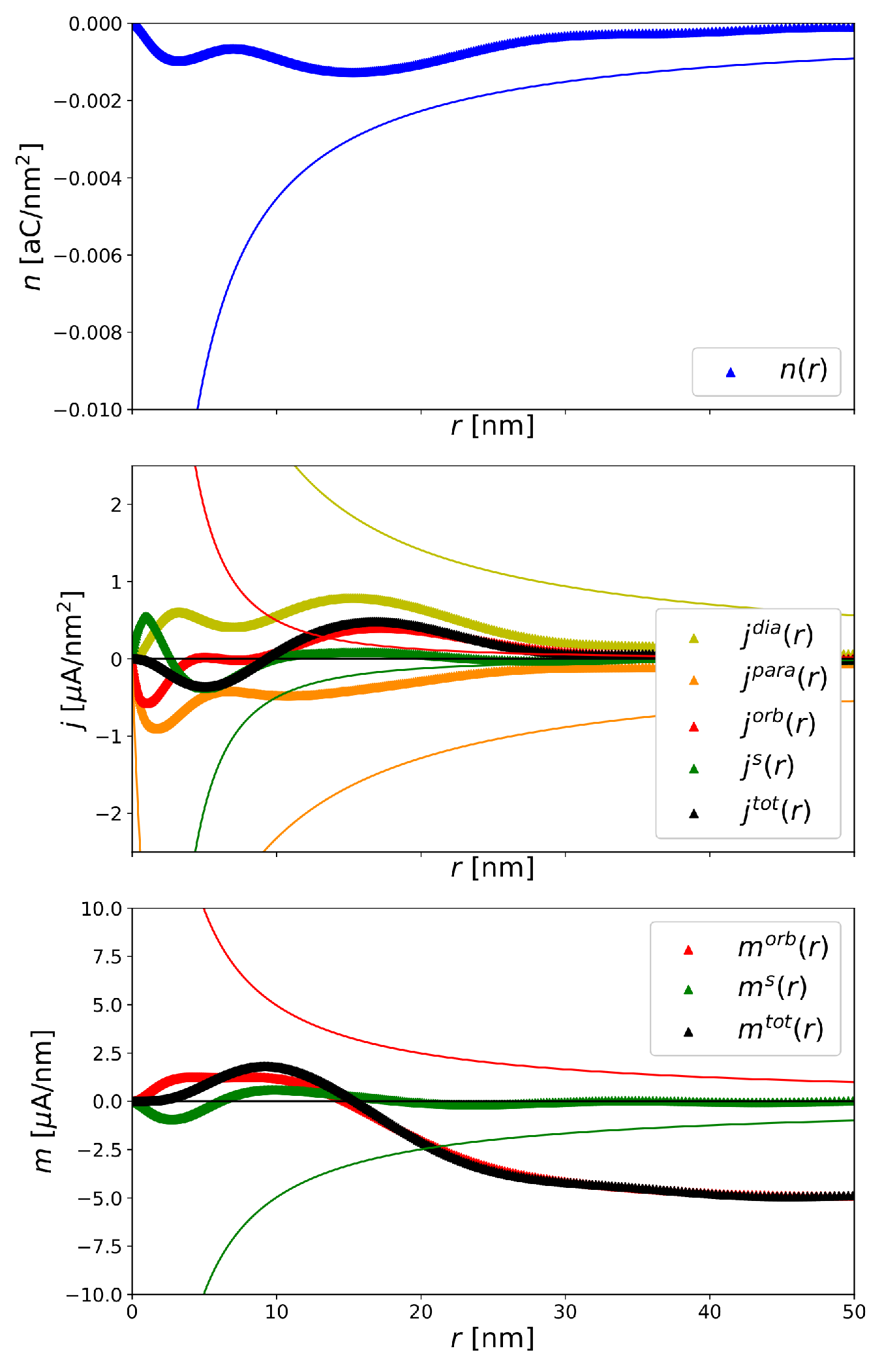}
     \end{subfigure}
     \hfill
        \caption{From top to bottom: Radially resolved charge, current and magnetization densities for $A_\phi=-0.186$ $\mu$Tm. 
        Thin continuous lines correspond to the analytic solution of the fully filled, infinitely degenerate lowest band at $E_F=0^+$. Solid lines, made of discrete triangles, correspond to the numerical solution for the fully filled flat band at $E_F=E_{2,1,1/2}=1.96$ meV. The later case nicely exemplifies the $B$-field induced Friedel oscillations for the charge density around the origin. They are accompanied by para- $j^{\rm para}$ and diamagnetic $j^{\rm dia}$, as well as magnetization current $j^{\rm s}$  oscillations, which result in a total current $j^{\rm tot}$ oscillating around the origin. Interestingly, the corresponding total magnetisation of this band indicates a mostly paramagnetic response to the applied magnetic field, whereas the lowest band does not respond at all, i.e. $j^{\rm tot}_0=m^{\rm tot}_0=0$, thanks to the exact cancellation of the orbital and spin contributions for $g_s=2$.  }
        \label{fig:denEF2}
\end{figure}

In a next step, we evaluate the density expressions given in Eqs.~(\ref{eq:chargensdef})-(\ref{eq:magn}) for the fully occupied, infinitely degenerate, lowest level at $E_{n,n,1/2}=0$. 
Surprisingly, the resulting infinite series converges to the following thermodynamic limit solution ($N\rightarrow\infty$) for the charge density in radial coordinates:
%

\begin{empheq}[box=\fbox]{align}
n_0(r,\phi)&=q\sum_{n=0}^\infty \psi_{n,n,\frac{1}{2}}^*\psi_{n,n,\frac{1}{2}}\nonumber\\
&=\frac{q^2 A_\phi}{\hbar \pi}\frac{ e^{-\frac{2 qA_\phi}{\hbar} r}\sinh(\frac{2 qA_\phi}{\hbar}
r)}{r},\label{eq:lowest_charge_dens}
\end{empheq}

where we applied the eigenstates explicitly given in Eq.~(\ref{eq:eigf_r_0}) of App.~\ref{app:ef} and used the series expansion of $\sinh(x)/x=\sum_{n=0}^\infty x^{2n}/(2n+1)! $. Notice, we can also find the corresponding simple closed form solution for $\tilde{n}_0(k)$ in reciprocal $k$-space (see App.~\ref{app:reciproc_dens}), which diverges (!) for $k\rightarrow 0$. 
Now, we utilize Eq. (\ref{eq:lowest_charge_dens})  to derive simple closed form solutions for the many-body current densities defined in Eqs. (\ref{eq:currdenspdef}) - (\ref{eq:currdenssdef}) in a similar fashion,
%
\begin{empheq}[box=\fbox]{align}
\vec{j}_{0}^{\rm para}(r,\phi)&= \frac{\hbar q}{2\pi m r}\sum_{n=0}^\infty  n \psi_{n,n,\frac{1}{2}}^*\psi_{n,n,\frac{1}{2}}\nonumber\\
&=\frac{q^3 A_\phi^2 }{\pi \hbar m}\frac{e^{-\frac{2 qA_\phi}{\hbar} r}}{ r}\cdot\nonumber\\
&\ \ \bigg(\cosh\Big(\frac{2 qA_\phi}{\hbar} r\Big)-\frac{\sinh\big(\frac{2 qA_\phi}{\hbar}r\big)}{\frac{2 qA_\phi}{\hbar} r}\bigg)\vec{e}_\phi\\
\vec{j}_{0}^{\rm dia}(r,\phi)&=
-\frac{q A_\phi}{ m}n_0(r,\phi)\vec{e}_\phi\\
\vec{m}_{0}^s(r,\phi)&=\frac{\hbar }{2 m} n_0(r,\phi)\vec{e}_z\\
\vec{j}_{0}^s(r,\phi)&=-\frac{d}{dr} m_z^s(r)=-\vec{j}_{0}^{\rm para}(r,\phi)-\vec{j}_{0}^{\rm dia}(r,\phi)\label{eq:magn_curr}
\end{empheq}
The last relation between orbital and magnetization current densities derived in Eq. (\ref{eq:magn_curr}) arises, when comparing the explicit results of $\vec{j}_{0}^{\rm para}$ and $\vec{j}_{0}^{\rm dia}$ with $\vec{j}_{0}^s$. This is a truly astonishing result, since this means that the total current density of the fully occupied lowest band vanishes exactly on the entire domain (!) of the infinitely extended 2D electron sheet (see Fig.~\ref{fig:denEF2} and \ref{fig:dos_angular}) %
\begin{eqnarray}
\vec{j}_{0}^{\rm tot}(r,\phi)&:=&\vec{j}_{0}^{\rm para}+\vec{j}_{0}^{\rm dia}+\vec{j}_{0}^s=0,
\end{eqnarray}
This automatically indicates a zero total magnetic response $\vec{m}^{\rm tot}=0$, where we used  the general definition for the magnetization density: $\vec{j}:=\vec{\nabla}\times \vec{m}$ and $\vec{\nabla}\cdot \vec{m}=0$ with normalizability condition. 
One would typically only expect such a vanishing magnetization density (found for $E_F=0^+$) in free space, but not in the presence of the induced, strongly inhomogeneous charge distribution, as given by Eq. (\ref{eq:lowest_charge_dens}).  Notice that tiny deviations of the exact cancellation arise from relativistic quantum fluctuations, which would introduce small corrections to $g_s=2$.\cite{van1987new} 
The observed subtle cancellation effect between orbital and magnetization currents can only emerge if the Zeeman interaction is included in our many-electron problem, whereas, for example, for $g_s=0$ the lowest flat level $\mu=0$ must not (!) exist, i.e. one would  observe a similar zero total current density at $E_F=O^+$, but originating from the zero occupancy at finite radii $r$ instead (absence of charges).
For this reason, the total charge densities strongly deviate between the $g_s=0$ and the $g_s=2$ solution, where only the later one shows a pronounced aggregation at the origin. This fundamental difference is rather surprising, give the close resemblance of the respective single electron solutions derived in Sec. \ref{sec:anasol}.
Consequently, a quantum effect (Zeeman interaction) fundamentally alters the (local) properties of our system.

Notice that on a first sight our lowest flat level $\nu=0$ closely resembles the Landau solution, for a homogeneous magnetic field (see App. \ref{app:landau}) applied to a non-interacting electron gas, which also predicts flat Landau levels with zero orbital and zero magnetization currents.  However, as already stated, in our case we find a highly inhomogeneous charge density distribution instead and the zero total magnetization is only reached thanks to opposing magnetization and orbital currents.  


\subsection{General Many-body Solution $(E_F\geq 0)$ \label{sec:intermediate}}
In a next step, we investigate the many-electron problem for $E_F>0$ numerically, in the vicinity of the magnetic field impurity  at $r=0$. Again, we assume fully filled bands throughout the  calculations. Fortunately, the numerical results reveal that we reach locally converged many-body solutions around the origin, with only a limited amount of eigenstates. This convergence in real space is rather surprising, since  we deal with an infinite amount of electrons, where infinite many energy level are infinitely degenerate. A problem that in principle cannot be represented on a computer. However, thanks to the exponential localisation of the states, simulations show that we can indeed reach numerically converged real-space, many-body solutions in the vicinity of the magnetic impurity.

Before continuing our computational analysis, we comment on the delicate choice of a reasonable parameter range for our investigations, which will hopefully become experimentally accessible in the near future. For this purpose, we try to minimize the magnetic flux through the 2D electron gas sample by choosing a small circular vector potential $A_\phi$. However, on the one hand, this comes at the cost of reducing the allowed energy domain given in Eq.~(\ref{eq:Ebound_domaintot}) and thermal noise may become an issue. On the other hand, reducing $A_\phi$ will also reduce the real space density of states for each energy level, as we can immediately infer from the scaling of the radial expectation value of a single electron 
\begin{eqnarray}
\langle \hat{r}\rangle_{nn1/2}
&=&\frac{\hbar}{q A_\phi }(n+1), \label{eq:r_n_zero}
\end{eqnarray}
at $E_F=0$. Notice that the derivation of Eq. (\ref{eq:r_n_zero}) is straightforward, since the associate Laguerre polynomial contribute trivially $L_{n-l}^{2l}=1$ to the involved eigenstates for $n=l$. Likewise scaling results hold for the density of states in higher lying bands, as we can infer from our numerical calculations. Consequently, with small $A_\phi$ a 2D electron gas with very low charge density may be required to investigate lower lying energy bands. Therefore, one has to ensure that Wigner crystallisation, i.e. a phase where the Coulomb interaction between the electrons dominates, does not hamper the results.\cite{smolenski2021signatures} Having made all these preliminary consideration, we suggest $A_\phi=-0.186$ $\mu$Tm 
as a reasonable choice, which will be used throughout our work. 
It ensures that the localised states in the lowest energy band at $E_F=0^+$ could in principle be populated solely, for a 2D material (e.g. transition metaldichalcogenide monolayers\cite{smolenski2021signatures}) with an extremely low electron density of $n_{2D}=10^{11}\ \rm cm^2$ and an effective mass $m^*\approx 1$, when considering a radial area defined by $r_{\rm max}=50$ nm. 
Such dilute electron gas have been realised experimentally, for which Wigner crystallisation could be avoided at temperatures above $T_W\approx 11\pm 1$  K (measured in absence of magnetic fields).\cite{smolenski2021signatures} Clearly, for higher electron densities, which implies $E_F>0$ in our setting, the Wigner crystallisation issue becomes less severe and lower temperatures could be reached to suppress thermal noise. 
Nevertheless, our selected $A_\phi$-value ensures that the energetic regime of the derived bound state solution is wider than the thermal noise, i.e. on the order of a $E_{A^2}\approx 3 \rm{meV}\sim T_A\lessapprox 35 \rm K $, which should in principle allow measurements down to $E_F=0^+$. Eventually, our considerations to minimize the magnetic field strength suggests the preparation of a (state of the art) dilute 2D electron gas within a temperature regime  $T_W\lessapprox T\lessapprox T_A$.
Notice that the homogeneous $B_{\rm hom}$-field equivalent, which generates the same magnetic flux  through a circular surface with $100$ nm diameter as provided by our inhomogeneous setting, is given by $B_{\rm hom}\approx 7.5$ T. A value that can routinely be achieved for homogeneous Landau settings.\cite{smolenski2021signatures} However, the experimental realisation of a constant $1/r$-decaying field shape, will require considerable experimental effort. Potential setups may facilitate magnetic lensing with meta materials or shaping the fields with (pumped) cavities, which offer a versatile approach to tailor electromagnetic-fields down to the nano-scale.\cite{hubener2021engineering}

\subsubsection{Magnetic Field Driven Charge and Current Density Oscillations}
\begin{figure*}
     \centering
     \begin{subfigure}[b]{0.45\textwidth}
         \caption{}
         \centering
         \includegraphics[width=\textwidth]{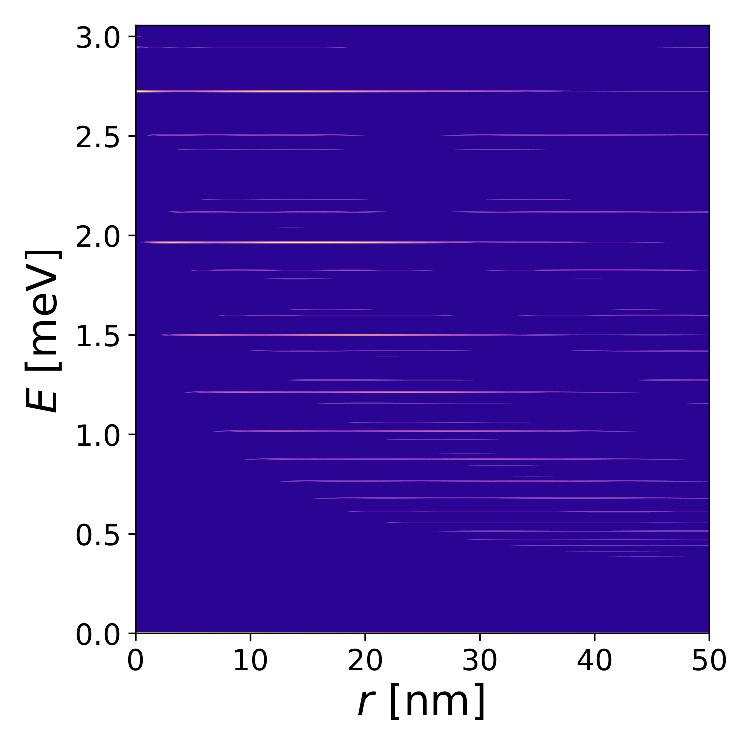}
         \label{fig:dosheat}
     \end{subfigure}
     \hfill
     \begin{subfigure}[b]{0.45\textwidth}
         \centering
         \caption{}
         \includegraphics[width=\textwidth]{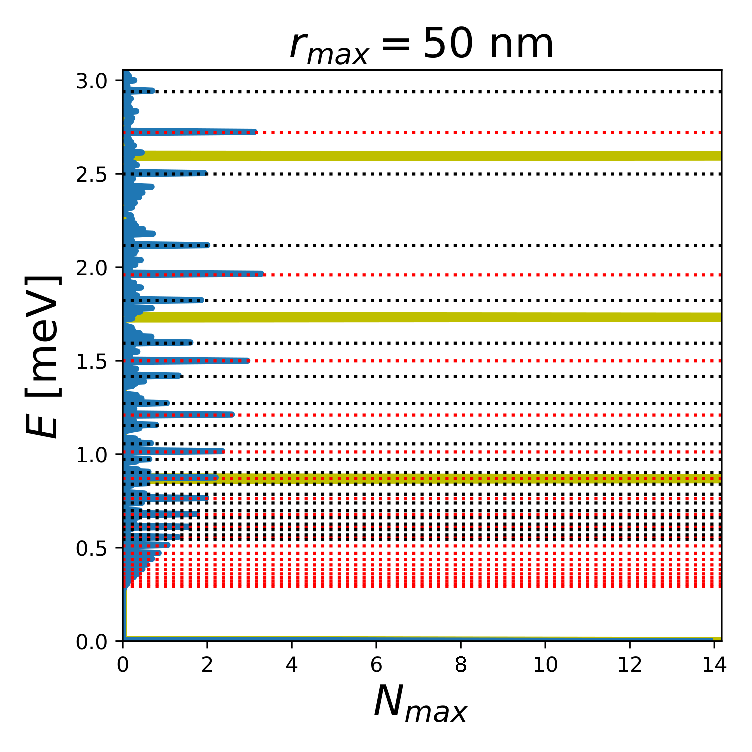}
         \label{fig:dos_N_landau}
     \end{subfigure}
        \caption{On the left: Radial density distributions reveal the quantization of the energy continuum around the origin, due to the applied inhomogeneous magnetic field with $A_\phi=-0.186$ $\mu$Tm. 
        On the right: The energy and density structure of the emergent levels is visualised based on the number of electrons $N_max$ contained within each level up to a radial distance of $r_{\rm max}=50$ nm. The horizontal yellow lines indicate the standard Landau-levels for a setup with identical magnetic flux, measured through the circular surface area limited by $r_{\rm max}.$ Dashed horizontal lines identify the most (red) and second  most (black) prominent levels, which are given by  $\nu=1$ and $\nu=2$. Notice that the observed DOS discretisation seems to be a local effect, which vanishes for larger radii $r_{max} \gg 50$ nm, where the discrete energy spectrum becomes denser (i.e. almost continuous). }
\end{figure*}

After having determined a reasonable field strength, we continue by investigating the radially resolved charge density with respect to the energy around the origin (see Fig. \ref{fig:denEF2}).
This analysis reveals a remarkable feature of our system. It depicts that a discrete, flat band-like, density structure emerges close to the origin. At a first sight, this appears to contradict our earlier definition of the distorted Landau levels. However, the here observed energetic quantisation of the density of state (DOS) has two fundamental limitations: First, it remains restricted to  the vicinity of the origin, whereas for larger radii the charge density becomes more and more continuous with respect to $E$. Second, the usual angular transition dipole selection rules $\Delta l=\pm 1$ effectively prevent significant inter-band transition between neighbouring DOS levels (e.g. see linear response Hall conductivity in Sec.~\ref{sec:hallcond}). Hence, the intriguing energy band structure in Fig.~\ref{fig:dos_N_landau} will determine the physics of our setup most likely only for very specific observables and perturbations, but not generally as it was the case for Landau levels.  

Nevertheless, the observed discrete DOS pattern 
has some interesting properties that we would like to mention and compare with the ubiquitous Landau levels. For example, the usual equidistant energy spacing is broken for our discrete DOS plateaus (see Fig.~\ref{fig:dos_N_landau}). Furthermore, in our case the DOS does not only depend non-trivially on the applied vector potential strength $A_\phi$, but also on the energy $E$ and the considered integrated surface around the origin ($r_{\rm max}$). For this reason, we do not have access to the scaling of the Fermi-energy with respect to the externally applied field throughout this work. 
As mentioned earlier, the DOS discretization pattern vanishes for large $r_{\rm max}$ due to the dense nature of the spectrum. 
This indicates that the physical origin of the discretization is most likely related to the decaying magnetic field and not a direct consequence of the constant vector potential $A_\phi$, which determines the asymptotic properties of our setting.
In more detail, we find that the most dominant density plateaus around the origin can be attributed to the quantum numbers $n,l=n-1,s=\frac{1}{2}$, which introduces a $n^{-1}$ decaying pattern in energy spacing, i.e.
\begin{eqnarray}
E^{\mathrm{tot}}_{n,n-1,\frac{1}{2}}&=&\frac{4q^2 A_\phi^2}{m} \cdot\frac{n}{(1+2n)^2},\ n\geq 1
\end{eqnarray}
 marked by the red dashed lines in Fig. \ref{fig:dos_N_landau}. Notice, this pattern exactly corresponds to the allowed intra-level transitions of our distorted Landau level at $\nu=1$, which contains the radially most localised electrons with $E>0$. Less dominant patterns are found with decreasing order for  $\nu>1$ (see black dashed lines for $\nu=2$ in Fig. \ref{fig:dos_N_landau}).
Another interesting aspect of the inhomogeneous field arises if one compares the density of states for our plateaus with the Landau solution  assuming an equal magnetic flux through the surface defined by $r_{\rm max}$ (yellow lines in  Fig. \ref{fig:dos_N_landau}). One immediately notices that the Landau levels have an increased DOS and the energetic spacing is considerably larger. Another major difference is that the number of electrons per Landau level scales quadratically with $r_{\rm max}$, whereas in our case the scaling is non-trivial except for the lowest band for which Eq. (\ref{eq:r_n_zero}) suggests a linear (!) scaling in $n_{\rm max}\propto \langle \hat{r}\rangle_{n_{\rm max}n_{\rm max}1/2}=r_{\rm max}$.

After having discussed this local DOS discretization, we focus next on the  charge and current-density observables, around the origin, which are computationally accessible for a finite number of one-electron states.

\begin{figure*}
     \centering
     \begin{subfigure}[b]{0.8\textwidth}
         \centering
         \includegraphics[width=\textwidth]{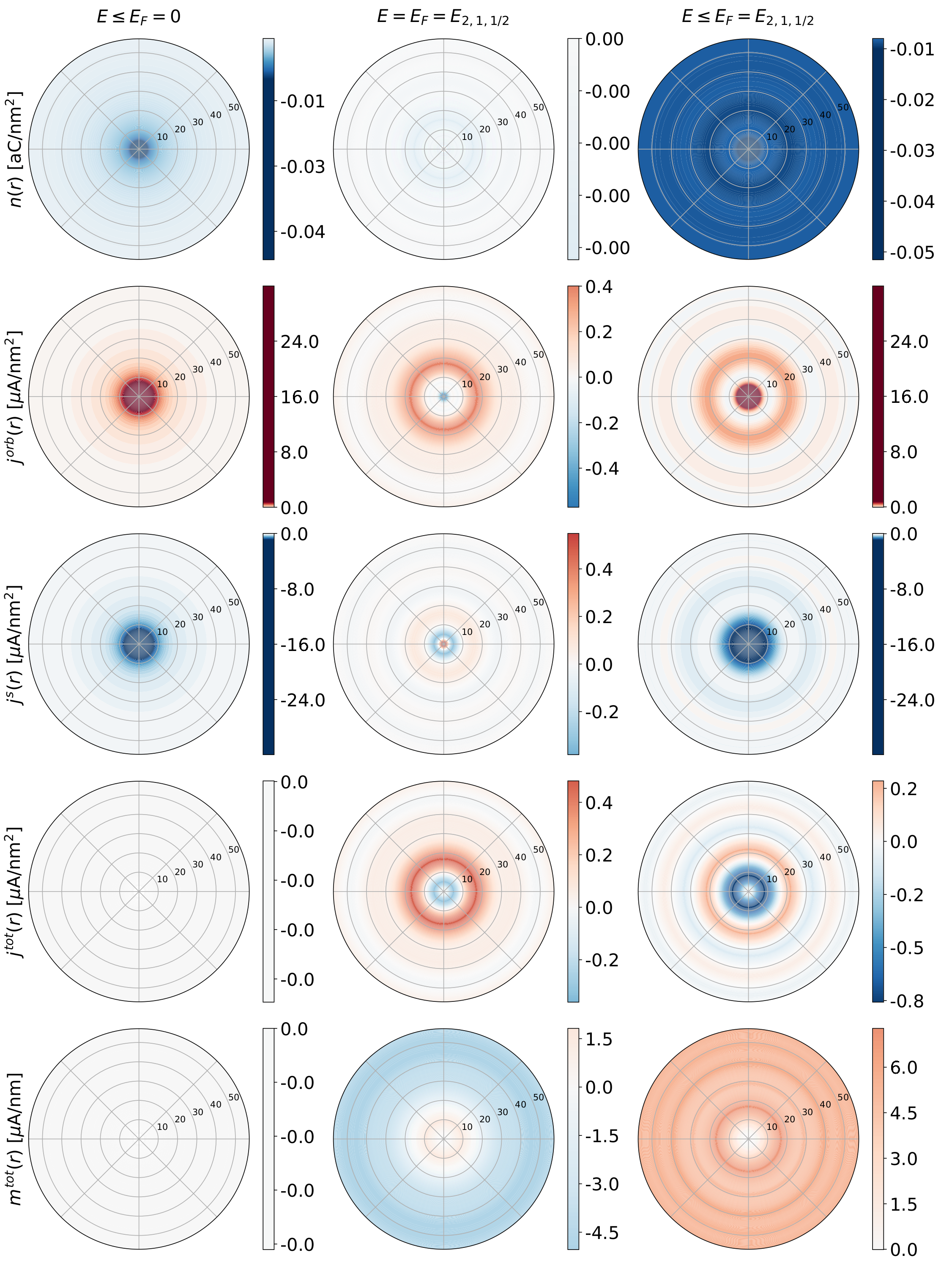}
     \end{subfigure}
        \caption{Angularly resolved density distribution for $A_\phi=-0.186$ $\mu$Tm. 
        The color-coding of each row is fixed to ensure horizontal comparability, whereas the displayed color-bars extend over the entire value range, which accounts for the substantial inhomogeneity at the origin. 
        Left column shows the analytic solution of the lowest band, orbital $j^{\rm orb}$ and magnetisation $j^{\rm s}$ currents cancel exactly and lead to a zero magnetic response $m^{\rm tot}=0$ to the applied $B$-field. The right column corresponds to the total densities of the system with occupied bands up to $E_F=E_{2,1,1/2}$. It demonstrates that the magnetic induced oscillatory behaviour persists for the entire system and is not only restricted to specific bands (e.g. at $E_F=E_{2,1,1/2}$ as displayed in the middle column). However, we find that the joint magnetic response of all filled levels always remains diamagnetic (e.g. $m^{\rm tot}\geq0 \ \forall r$ at the bottom of the right column), whereas selected degenerated energy levels may respond paramagnetically (see bottom figure in the middle column).  }
        \label{fig:dos_angular}
\end{figure*}

In Fig. \ref{fig:denEF2} charge, current and magnetization density profiles are displayed with respect to the radial distance $r$ from the origin. Thin lines correspond to the previously derived analytic results for the fully filled band (energy levels) at $E_F=0$, whereas bold triangles indicate a  numerical solution  for a prototypical, fully filled, infinitely degenerated energy level at $E_F=E_{2,1,1/2}$. One immediately notices that the field inhomogeneity introduces a magnetic defect in the electron gas, which gives rise to radial density fluctuations, except for the lowest lying flat band (top panel of Fig.~\ref{fig:denEF2}). The fluctuations of the charge density in Fig.~\ref{fig:denEF2} resemble Friedel oscillations, which typically emerge in the vicinity of charge impurities. However, for our magnetic-field impurity we additionally find non-vanishing, circular charge currents (in $\phi$-direction) for $E_F>0$, which also show oscillatory behaviour in radial direction (middle panel in Fig. \ref{fig:denEF2}) with corresponding  magnetization fluctuations in the bottom panel of Fig. \ref{fig:denEF2} ($\vec{j}(\vec{r})=\vec{\nabla}\times\vec{m}$, with $\vec{\nabla}\cdot \vec{m}=0$).  Interestingly, the radially resolved total current density $j^{\rm tot}$ seems to oscillate around zero, i.e. they can change direction. This is a true quantum effect, which could not emerge for identical classical charges, subject to the inhomogeneous magnetic field along $z$. In more detail, one can show that every single electron current expectation value  is positive $\langle j^{\rm tot}\rangle_{n,l,s}>0$ for our setting, i.e. the diamagnetic term dominates (see Eq. (\ref{eq:totcur}) in App. \ref{app:current_magn}). This automatically implies that the total magnetic response of our system will be diamagnetic, which agrees with the Landau case.\cite{landau1930diamagnetismus} However, in the vicinity of the magnetic impurity things can change fundamentally. 
While overall the charge and current density fluctuations remain qualitatively similar, i.e. independent of considering all bands up to the Fermi level (right column in Fig.~\ref{fig:dos_angular}) or only the highest occupied band (middle column in Fig.~\ref{fig:dos_angular}), things change, when investigating the magnetization density. In that case, one observes that the magnetic response can become paramagnetic for certain bands  (see bottom row in Fig.~\ref{fig:dos_angular}), while the total magnetic response always remains diamagnetic.  A crucial ingredient for this effect is the proper consideration of the Zeeman interaction as well as spin-dependent magnetization currents, in order to achieve the paramagnetic response of certain (degenerate) energy bands. 
 
To summarize our many-body results up to this point, we would like to mention that overall the fundamental driving mechanism investigated so far is mostly local, and is mainly be related to the strong magnetic field inhomogeneity, which decays as $1/r$. In other words, the constant circular vector potential $A_\phi$ plays only a minor role for the observed  local density  aggregation and fluctuations or for the emergence of the DOS plateaus. However, things change for different observable, as we will see next.



\subsubsection{Vector Potential Driven Sign Flip in Hall Conductance 
\label{sec:hallcond}}

In a next step, we investigate the radially resolved $r_x$ Hall conductivity tensor $\sigma_{xy}$ for a static, homogeneous electric field perturbation in $y$-direction. At zero temperature, the linear response Hall conductivity assumes the following simple form,\cite{thouless1982quantized,kohmoto1985topological}
\begin{eqnarray}
   \sigma_{xy}(r_x,E_F)&=&\frac{\iu e^2}{\hbar}\sum_{E_a<E_F<E_b} \frac{1}{(E_a-E_b)^2}\cdot\nonumber\\
   &&\Big[\bra{a}\delta(r_x-r)\hat{v}_x\ket{b}\bra{b}\hat{v}_y\ket{a}-\nonumber\\
   &&\ \ \bra{a}\hat{v}_y\ket{b}\bra{b}\delta(r_x-r)\hat{v}_x\ket{a}\Big],\label{eq:Halldef}
\end{eqnarray}
%
 The velocity operator $\hat{\vec{v}}=(\hat{\vec{p}}- q \vec{A})/m$ can be written as 
\begin{eqnarray}
        \begin{bmatrix}
         \hat{v}_x \\
         \hat{v}_y
         \end{bmatrix}
    &=&\frac{1}{m}\bigg(-\iu\hbar
    \begin{bmatrix}
         \cos\phi \partial_r-\frac{\sin\phi}{r}\partial_\phi \\
         \sin\phi \partial_r+\frac{\cos\phi}{r}\partial_\phi 
         \end{bmatrix}
        -qA
        \begin{bmatrix}
         -\sin\phi \\
         \cos\phi
         \end{bmatrix}
         \bigg)
\end{eqnarray}
where the Cartesian components $x,y$ are expressed in radial coordinates. 
The radially resolved Hall conductivity in Eq. (\ref{eq:Halldef}) has two major advantages compared with the standard integrated quantity. First, it reveals the rich local conductivity variations, which we anticipate due to the observed charge and current density oscillations around the origin. Second, thanks to the exponential localisation of the single electron eigenstates,  we can indeed determine locally converged Hall conductivities at the magnetic impurity for infinite system sizes, which can be utilized to infer asymptotic properties of our system. 
Notice that the involved angular and spin integrals are solved for the  transition velocity elements, which give rise to the same angular and spin selection rules  as previously seen in  Eq.~(\ref{eq:diprulerig}) for the transition dipole moments.
This reduces the computational demand of the summation over occupied $\ket{a}$ and unoccupied states $\ket{b}$ considerably, since the only non-vanishing contributions arise from $l_a=l_b\pm 1$  (see App.~\ref{app:ef} for more details on the numerics). 

\begin{figure*}
     \centering
     \begin{subfigure}{0.45\textwidth}
         \caption{}
        \centering
         \includegraphics[width=84.5mm]{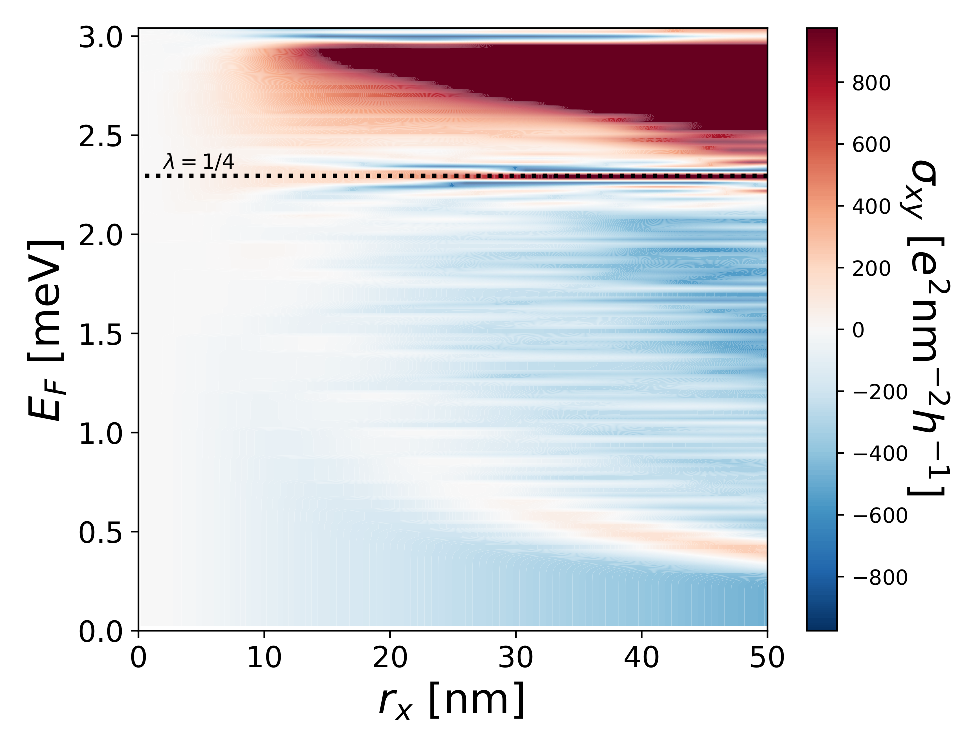}
        \label{fig:hallcondloc}
     \end{subfigure}%
     \begin{subfigure}{0.45\textwidth}
         \caption{}
         \centering
         \includegraphics[width=65mm]{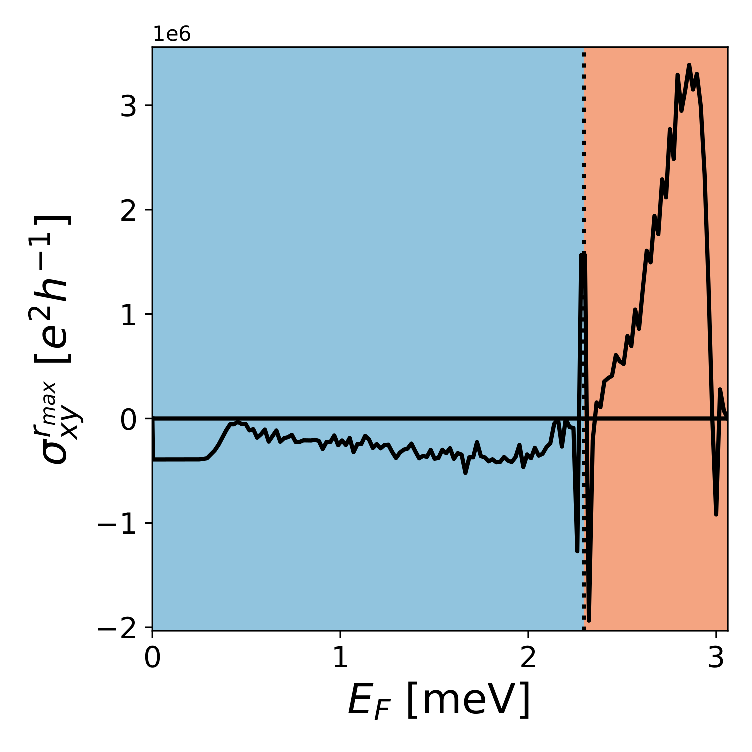}
         \label{fig:hallcondint}
     \end{subfigure}
        \caption{(a) Radially resolved Hall conductivity $\sigma_{xy}$ for a homogeneous electric field perturbation along the $y$-direction. To reach a locally converged solution within $r_{max}=50$ nm, a large number of single electron eigenstates with $\langle r\rangle\gg r_{max}$ is required, which indicates that the locally observed Hall conductivity switch is driven by the constant anisotropic vector potential instead of the localised inhomogeneous magnetic field.  (b) Consequently, the integrated Hall conductivity transition at $E_F=E_{A^2}(1-1/4)$ is expected to persist in the asymptotic limit $r_{max}\rightarrow\infty$, which automatically implies that spin contributions are of minor importance for this observable, except for $E_F=0+$. }
\end{figure*}

The locally converged Hall conductivity $\sigma_{xy}(r_x, E_F)$  is displayed in Fig.~\ref{fig:hallcondloc} with respect to $r_x$ (angularly integrated in $\phi_x$) up to $r_{\rm max}=50$ nm for the static, homogeneous electric field perturbation along $y$. 
One immediately notices that the magnetic inhomogeneity leads to a depletion of the conductivity close to the origin ($r_x<10$ nm), i.e. a whitening of the color pattern, which is caused by the relatively low electron density of the excited states in combination with large $\Delta E=E_b-E_a$ for the allowed transitions.
However, our locally converged Hall conductivity suggests a remarkable asymptotic feature for our setting: the emergence of quantum-Hall phases where the sign of the conductance fluctuates. In more detail, we find that the local Hall conductivity is negative (blue)  for $E_F<E_\lambda$, at $\lambda=1/4$ with $E_\lambda$ given in Eq.~(\ref{eq:elambda}), and positive for higher lying Fermi energies (red). This sign change appears even more pronounced for the integrated conductivity pattern $\sigma^{r_{\rm max}}_{xy}(E_F):=\int_0^{r_{\rm max}}\sigma_{xy}(r_x,E_F)r_x dr_x $ shown in Fig.~\ref{fig:hallcondint}. It is important to mention that this phenomenon of the sign change in the Hall conductance also shows up experimentally for the Hofstadter butterfly in Moir\'{e} materials,~\cite{DeanButterfly, WangButterfly} for which the homogeneity of the system is broken by the lattice periodicity, whereas in our case we rely on a $B$-field inhomogeneity.

To reach our numerical results, it turns out that the accurate consideration of a relatively large number of single electron states is vital to reach converged results and in particular no reduction of the allowed $\Delta n$ transitions can be applied to speedup the calculations (caused by the $1/\Delta E^2$-dependency of $\sigma_{xy})$. The convergences becomes particularly tricky for larger Fermi energies, due to previously discussed increase in the delocalisation of states, i.e. the sharp conductivity drop (below zero) observed around $E_F=3$ meV is likely to be a numerical artifact. 
The relevance of a large amount of states with $\langle r\rangle\gg r_{\rm max}$ indicates that our observation is mainly driven by the constant $A_{\phi}$ vector potential and not by the $1/r$-decaying $B$-field, significant solely in the vicinity of the origin. This automatically suggests that the Zeeman interaction should not play a significant role, which can indeed be verified numerically (see almost equivalent results for $g_s=0$ displayed in Figure S1 of the Supporting Information).
Clearly, from our locally converged solution we can only infer asymptotic properties, and considerable future research effort will be required to further validate our results theoretically as well as experimentally. However, the fundamental origin of the two different Hall conductivity phases is likely to be a consequence of the distorted Landau level structure identified in Figs.~\ref{fig:denEF0} and \ref{fig:transitions}, which possess a clearly positive curvature in the positive Hall conductivity phase and \textit{vice versa} for the negative phase.  

\begin{figure*}
     \centering
     \begin{subfigure}{0.45\textwidth}
         \caption{}
        \centering
         \includegraphics[width=84.5mm]{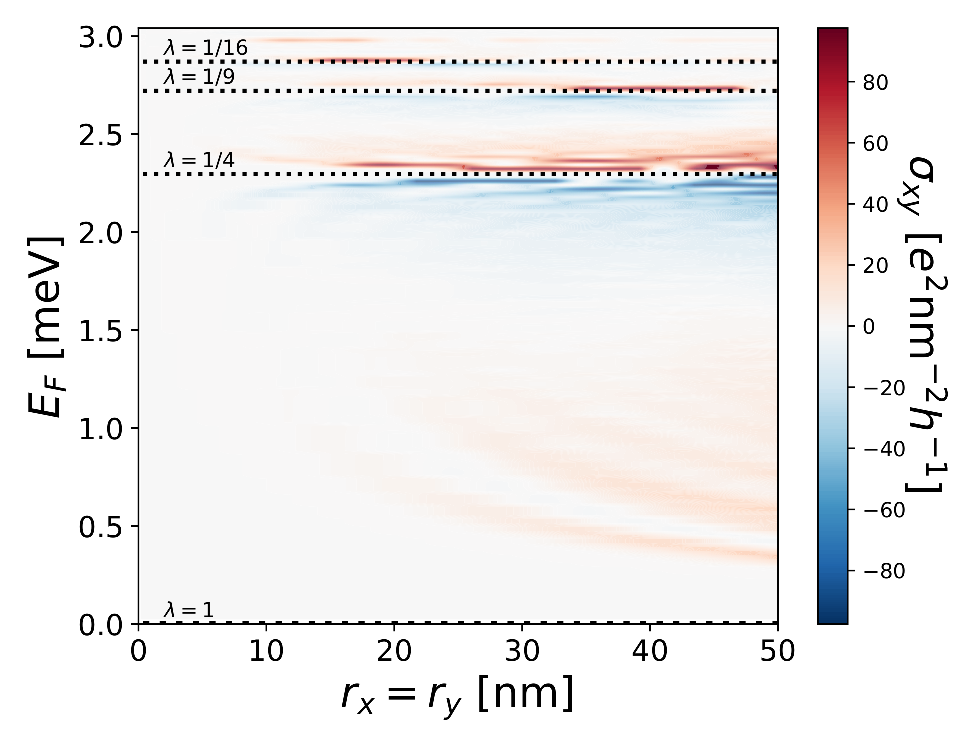}
        \label{fig:hallcondloccyl}
     \end{subfigure}%
     \begin{subfigure}{0.45\textwidth}
         \caption{}
         \centering
         \includegraphics[width=65mm]{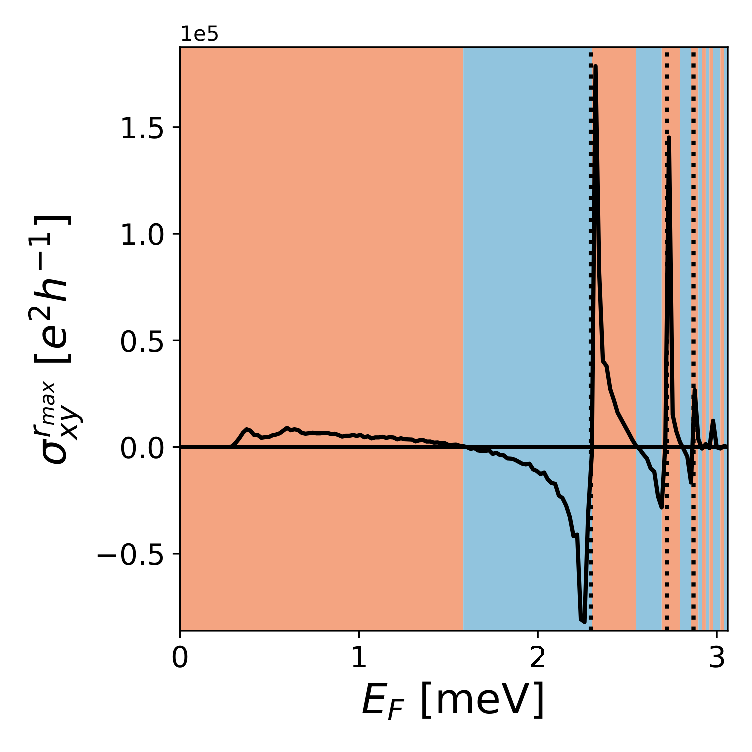}
         \label{fig:hallcondintcyl}
     \end{subfigure}
        \caption{(a) Radially resolved Hall conductivity $\sigma_{xy}$ measured at $r_x$ for a cylindrical  electric field perturbation at $r_x$ along $y$.  Multiple sharp conductivity transitions (dotted horizontal lines) are observed for this radially localised (!) perturbation, which follow  a Rydberg series (labeled by $\lambda$) that alternates with rather smooth transition of opposite sign. (b) Integrated Hall conductivity transition pattern (summation over many local measurements at different radii), which clearly allows to distinguish smooth from sharp transitions with respect to the Fermi Energy $E_F$.}
\end{figure*}

Moreover,  the rich spatial as well as energetic structure (e.g. density variation, distorted Landau levels) provides numerous opportunities to discover novel physical effects emerging for different types of perturbations (e.g. spatially or time resolved). 
Here, we exemplify our claim  for a specific, locally resolved Hall conductivity measurement, which reveals particularly interesting properties.
We assume our system is perturbed with a static constant electric fields directed in $y$-direction, which acts solely on the radial shells located at $r_y$. The resulting, $\phi_x$-integrated Hall conductivity in $x$-direction is measured at the position of the perturbation i.e. at $r_x=r_y$:
\begin{eqnarray}
   \sigma_{xy}(r_x,r_x,E_F)&=&\frac{\iu e^2}{\hbar}\sum_{E_a<E_F<E_b} \frac{1}{(E_a-E_b)^2}\cdot\\
   &&\Big[\bra{a}\delta(r_x-r)\hat{v}_x\ket{b}\bra{b}\delta(r_x-r)\hat{v}_y\ket{a}-\nonumber\\
   &&\ \ \bra{a}\delta(r_x-r)\hat{v}_y\ket{b}\bra{b}\delta(r_x-r)\hat{v}_x\ket{a}\Big].\nonumber\label{eq:Halldefloc}
\end{eqnarray}
The resulting Hall conductivity is displayed with respect to the Fermi level $E_F$ in Figs.~\ref{fig:hallcondloccyl} and \ref{fig:hallcondintcyl}. 
In contrast to the previous homogeneous perturbation, we find a fractional quantum Hall conductivity pattern with sharp (!) transitions that alternate with smooth sign changes in-between.
Interestingly, the sharp Hall conductivity transitions follow exactly the Rydberg series of the Hydrogen energy levels (horizontally dotted lines), i.e. they are observed at
\begin{eqnarray}
    E_\lambda&:=&E_{A^2}(1-\lambda)\label{eq:elambda}\\
    \lambda&=&\frac{1}{n_\lambda^2},\ n_\lambda\in\mathbb{N}.
\end{eqnarray}
Hence,  some Hydrogen properties are recovered at least for this specific perturbation, which one probably might have expected, due to the similarity of the corresponding partial differential equations.
Notice that similarly to the previous computations, the convergence of the Hall conductivity becomes increasingly complex at high Fermi Energies, i.e. for $\lambda\lessapprox 1/16 $, due to the strong delocalization of the wavefunctions.

\section{Conclusion and Outlook}

To our knowledge, this work establishes the first simple and explicit, analytical solution for an extended 2D electron gas subject to a static inhomogeneous magnetic field including the Zeeman interaction.
The resulting exact eigenstates provide access to the many-body properties of a non-interacting electron gas, which can be calculated numerically and even analytically (for $E_F=0^+$) in the thermodynamic limit. 
Based on those results, distorted Landau levels could be identified, which eventually give rise to spin-dependent, localised density / current oscillations as well as distinct switching between different asymptotic Hall conductivity phases, driven by the anisotropic vector potential instead of the decaying magnetic field.
Overall our findings highlight that our exact solution gives rise to a variety of fundamental new physical effects, which strongly deviate from the  Landau solution, locally as well as in the asymptotic limit.

However, certainly the experimental verification of our theoretical results will require considerable future research effort. Nevertheless, we believe this will be a highly rewarding endeavor.
Despite our fundamental observations made so far, we are far from having explored the full potential of our solution yet. Indeed, we believe that our exact solution opens the door to enter  novel physical regimes providing numerous theoretical and experimental opportunities, which are awaiting to be explored. 
For example, novel effects are anticipated to emerge at a zero Hall conductivity phase transition. Moreover, at the moment, we still lack an asymptotic description of the density of states, which would allow to determine the dependency of the Fermi-energy $E_F(A_\phi)$ with respect to the applied vector potential strength for a fixed electron density of the underlying 2D material. Having access to these asymptotic properties, could enable the  exploration of the magnetic susceptibility (e.g. De-Haas van Alphen like effects\cite{landau1930diamagnetismus}) or the emergence of different Shubnikov-De Haas\cite{schubnikow1930magnetische} like conductivity oscillations for varying magnetic field strengths. Generally speaking, the application of Kubo's linear response theory can be extended to further (static, localised, time- and even spin-dependent) perturbations, beyond the measurement of the Hall conductivity. Those theoretical investigations are ideal to propose and design novel experimental setups. From a theoretical perspective the formal connection to established impurity models, such as Anderson localisation\cite{anderson1958absence} or the Kondo effect\cite{kondo1964resistance}, is still absent. In contrast to our setting, they are formulated in reciprocal space assuming periodic systems.  Introducing periodicity  (e.g. due to the presence of a lattice) may not be trivial in our setting, due to the non-perturbative nature of the magnetic field induced impurity. Potentially, one could either try to introduce periodicity perturbatively, or one could reach for  a Hofstadter-butterfly type of effect,\cite{hofstadter1976energy} assuming periodic magnetic-field perturbations instead. 
Apart from raising these open theoretical questions, our analytic solution may also help the numerical description of light-(quantum)matter interactions in inhomogeneous magnetic fields.  For example, the discovered analytic eigenfunctions could be a reasonable basis-set choice for numerical simulations of differently decaying, radial symmetric $B$-fields, which may be easier accessible experimentally than our $1/r$-solution. Furthermore, corrections from electron-electron (Coulomb) or even current-current (transverse) interactions should be straightforward to include numerically on different levels of approximations (e.g. Jellium setting). On the long run, it would also be exciting to investigate our inhomogeneous setting in the context of (doped) two-dimensional heterostructures, similar to Landau levels physics in 2D Moir\'e materials,\cite{andrei2021marvels} which are governed by the interplay of topological, correlation as well as band structure effects.
Overall, we believe that the discovered analytic solution will serve as a paradigmatic model for a large number of future theoretical as well as experimental investigations.

\begin{acknowledgments}
We thank Simone Latini for inspiring discussions.
This work was made possible through the support of the RouTe Project (13N14839), financed by the Federal Ministry of Education and Research (Bundesministerium für Bildung und Forschung (BMBF)) and supported by the European Research Council (ERC-2015-AdG694097), the Cluster of Excellence “CUI: Advanced Imaging of Matter” of the Deutsche Forschungsgemeinschaft (DFG), EXC 2056, project ID 390715994 and the Grupos Consolidados (IT1249-19). V.R. acknowledges support from the NSF through a grant for ITAMP at Harvard University.
The Flatiron Institute is a division of the Simons Foundation.
\end{acknowledgments}

\section*{Author Contributions}
D.S. initiated the project, discovered the simple closed form solution and performed corresponding analytic as well as numerical calculations. M.R. contributed to the mathematical accuracy and rigorosity. V.R. added expertise and calculations to connect with the homogeneous Landau setting.
All authors developed the physical interpretation and wrote the manuscript.

\section*{Data Availability Statement}
Numerical data available upon request.

\bibliographystyle{unsrt}
\bibliography{references}

\appendix


\section{Scaled Eigenfunctions\label{app:ef}}

To calculate radial expectation values, it can be convenient to express the eigenfunctions in Eq. (\ref{eq:eigf}) in terms of $r$: 
\begin{eqnarray}
\Psi_{n,l,s}&=&\frac{1}{\sqrt{\tilde{N}_{n,l,s}}}e^{il\phi}e^{-\frac{q A_\phi}{\hbar}\frac{2(l+s)}{2n+1}r}\nonumber\bigg(\frac{2q A_\phi}{\hbar}\frac{2(l+s)}{2n+1}r\bigg)^l\\
&&\cdot L_{n-l}^{2l}\bigg(\frac{2q A_\phi}{\hbar}\frac{2(l+s)}{2n+1}r\bigg)\chi(s).\label{eq:eigfr}
\end{eqnarray}
The corresponding normalisation constant changes to,
\begin{eqnarray}
\tilde{N}_{n,l,s}
&=&2\pi \bigg(\frac{2q A_\phi}{\hbar}\frac{2(l+s)}{2n+1}\bigg)^{-2}\frac{(n+l)!}{(n-l)!}(2n+1).
\end{eqnarray}
Notice the simple expression for the wave-functions at $E_F=0$, 
\begin{eqnarray}
\psi_{n,n,\frac{1}{2}}(\rho,\phi)&=&\frac{1}{\sqrt{2\pi(2n+1)!}}e^{\iu n\phi}e^{-\frac{\rho}{2}}\rho^n\chi(\frac{1}{2})\\
\psi_{n,n,\frac{1}{2}}(r,\phi)&=&\sqrt{\frac{2q^2 A_\phi^2}{\hbar^2\pi(2n+1)!}}e^{\iu n\phi}e^{-\frac{q A_\phi}{\hbar}r}\Big(\frac{2q A_\phi}{\hbar}r\Big)^n\nonumber\\
&&\cdot\chi(\frac{1}{2})\label{eq:eigf_r_0}
\end{eqnarray}
with $\rho=\frac{2q A_\phi}{\hbar} r$, which allows to perform the infinite summation over this infinitely degenerated many-body state. Notice that the condition $n=l$ leads to trivial associated Laguerre polynomials $L_0^{2l}=1$.

The electronic dipole transition selection rules in Eqs. (\ref{eq:diprulerig}) \& (\ref{eq:diprule}), of the single electron eigenfunctions given in Eq. (\ref{eq:eigf}) for $\rho\mapsto r$, arise from,
\begin{eqnarray}
\bra{n,l,s}\vec{r}\ket{n^\prime,l^\prime,s^\prime}&=&\bra{R_{n,l,s}} r\ket{R_{n^\prime,l^\prime,s^\prime}}\nonumber\\
&&\cdot|\bra{\Phi_{l}}\cos(\phi)\vec{e}_r-\sin(\phi)\vec{e}_\phi\ket{\Phi_{l^\prime}}|\bra{\chi_s}\ket{\chi_{s^\prime}}\nonumber\\
 &=&\int_0^\infty\frac{1}{\sqrt{\tilde{N}_{n,l,s}\tilde{N}_{n^\prime,l^\prime,s^\prime}}}e^{-\frac{\rho_{n,l,s}(r)+\rho_{n^\prime,l^\prime,s^\prime}(r)}{2}}\label{eq:radial_diptrans}\\
 &&\cdot\big(\rho_{n,l,s}(r)\big)^l\big(\rho_{n^\prime,l^\prime,s^\prime}(r)\big)^{l^\prime}\cdot L_{n-l}^{2l}(\rho_{n,l,s}(r))\nonumber\\
 &&L_{n^\prime-l^\prime}^{2l^\prime}(\rho_{n^\prime,l^\prime,s^\prime}(r))r^2 dr\delta_{l\pm1,l^\prime}\delta_{s,s^\prime}\nonumber\\
 &\propto& \frac{1}{A_\phi}
\end{eqnarray}
where the $A_\phi$-proportionality can be shown by suitable change of variable $x=\frac{\rho_{n,l,s}(r)+\rho_{n^\prime,l^\prime,s^\prime}(r)}{2}$, which eventually removes the $A_\phi$ from the exponential and Laguerre functions. Thus $A_\phi$ appears only as a prefactor of the integral. Accurate and efficient numerical evaluation of Eq. (\ref{eq:radial_diptrans}) can be performed in $x$-space by means of generalized Gauss-Laguerre quadrature. The radial parts of the subsequent velocity matrix elements  can be integrated in a similar fashion for the Hall conductivity in Eq. (\ref{eq:Halldef}),
\begin{eqnarray}
\bra{n,l,s}\hat{v}_x\ket{n^\prime,l^\prime,s^\prime}&=&
\bigg[\frac{-\iu\hbar}{m}\big(\bra{\Phi} \cos\phi\ket{\Phi^\prime}\bra{R} \partial_r\ket{R^\prime}\\
&&-\bra{\Phi} \sin\phi\partial_\phi\ket{\Phi^\prime}\bra{R} \frac{1}{r}\ket{R^\prime}\big)\nonumber\\
&&+\frac{qA_\phi}{m}\bra{\Phi} \sin\phi\ket{\Phi^\prime}\bra{R}\ket{R^\prime}\bigg]\bra{\chi_s}\ket{\chi_{s^\prime}}\nonumber\\
\bra{n,l,s}\hat{v}_y\ket{n^\prime,l^\prime,s^\prime}&=&
\bigg[\frac{-\iu\hbar}{m}\big(\bra{\Phi} \sin\phi\ket{\Phi^\prime}\bra{R} \partial_r\ket{R^\prime}\\
&&+\bra{\Phi} \cos\phi\partial_\phi\ket{\Phi^\prime}\bra{R} \frac{1}{r}\ket{R^\prime}\big)\nonumber\\
&&-\frac{qA_\phi}{m}\bra{\Phi} \sin\phi\ket{\Phi^\prime}\bra{R}\ket{R^\prime}\bigg]\bra{\chi_s}\ket{\chi_{s^\prime}}.\nonumber
%
\end{eqnarray}
The selection rules $\Delta l=\pm 1$ apply to all angular transition matrix elements (with different coefficients) and for the spin we find $\Delta s=0$. However, no exact selection rule applies to the radial transition matrix elements, in particular $\bra{R}\ket{R^\prime}\neq0$.

\section{Single Electron Currents and Magnetization\label{app:current_magn}}

Interestingly,  for our system simple closed form solutions can be calculated for the expected single electron currents:
\begin{eqnarray}
\vec{J}_{n,l,s}^{\mathrm{para}}&=&\frac{\hbar q}{2m\iu}\Big(\bra{n,l,s}\overrightarrow{\nabla}\ket{n,l,s}-\bra{n,l,s}\overleftarrow{\nabla}\ket{n,l,s}\Big)\nonumber
\\
&=& \frac{\hbar q}{m\iu}\bra{n,l,s}\frac{1}{r}\frac{\partial}{\partial \phi}\ket{n,l,s}\vec{e}_\phi\label{eq:paracur_phi}\\
&=& \frac{ q^2 A_\phi}{m}\frac{4l(l+s)}{(2n+1)^2}\vec{e}_\phi\rightarrow
\begin{cases}
0 &\text{if $E^{\mathrm{tot}}_{n,l,s}\approx E_{A^2}$}\\
\frac{ q^2 A_\phi}{m}\frac{2n}{2n+1}\vec{e}_\phi &\text{if $E^{\mathrm{tot}}_{n,n,\frac{1}{2}}= 0$}
\end{cases},\nonumber\label{eq:paracur}\\
\vec{J}_{n,l,s}^{\mathrm{dia}}&=&-\frac{q^2}{m}\bra{n,l,s}\vec{A}\ket{n,l,s}=-\frac{ q^2 A_\phi}{m}\vec{e}_\phi,\label{eq:diacur}\\
\vec{J}_{n,l,s}^{\mathrm{s}}&=&\int_{0}^{2\pi}\int_0^\infty (\vec{\nabla}\times \vec{m}^s)r dr d\phi\nonumber\\
%
&=& -\frac{2\hbar q}{m} s \bra{n,l,s}\frac{\partial}{\partial r}\ket{n,l,s} \vec{e}_\phi\label{eq:magncurphi}\nonumber \\
&\overset{P.I}=&\frac{\hbar q}{m}s \bra{n,l,s}\frac{1}{ r}\ket{n,l,s} \vec{e}_\phi\\
&=&\frac{4 q^2 A_\phi}{m}\frac{s(l+s)}{(2n+1)^2}\vec{e}_\phi
\rightarrow
\begin{cases}
0 &\text{if $E^{\mathrm{tot}}_{n,l,s}\approx E_{A^2}$}\\
\frac{ q^2 A_\phi}{m}\frac{1}{2n+1}\vec{e}_\phi &\text{if $E^{\mathrm{tot}}_{n,n,\frac{1}{2}}= 0$}
\end{cases},\nonumber\label{eq:magncur}
\end{eqnarray}
with
\begin{eqnarray}
\vec{J}_{n,l,s}^{\mathrm{tot}}:=\vec{J}_{n,l,s}^{\mathrm{para}}+\vec{J}_{n,l,s}^{\mathrm{dia}}+\vec{J}_{n,l,s}^{\mathrm{s}}\geq 0 \rightarrow
\begin{cases}
\vec{J}_{n,l,s}^{\mathrm{dia}} &\text{if $E^{\mathrm{tot}}_{n,l,s}\approx E_{A^2}$}\\
0 &\text{if $E^{\mathrm{tot}}_{n,n,\frac{1}{2}}= 0$}
\end{cases}.\label{eq:totcur}\nonumber\\
\end{eqnarray}
in state $n,l,s$. Single electron currents are visualised in Fig. S2 of the Supporting Information with respect to their radial expectation values and energy.
To obtain Eq. (\ref{eq:paracur_phi}), it was used that the radial part $R(r)$ of our wave function is real as well as the gradient operator in $\vec{e}_r$, which cancels the paramagnetic current in $r$-direction. With a similar argument only the $\vec{e}_\phi$ component of the applied curl operator survives for the magnetization current in Eq. (\ref{eq:magncurphi}).
For the solution of the integrals in Eq. (\ref{eq:paracur_phi}) the following relation was used,\cite{arfken1999mathematical}
\begin{eqnarray}
\bra{n,l,s}\frac{1}{r}\frac{\partial}{\partial\phi}\ket{n,l,s}&=&\frac{2q A_\phi }{\hbar N_{n,l}}\frac{2(l+s)}{2n+1}(\iu l)\cdot\\
&&\int_0^{2\pi}\int_0^\infty e^{-\rho} \rho^{2l} \big(L_{n-l}^{2l})^2 d\rho d\phi\nonumber\\
&=& \iu\frac{2q A_\phi }{\hbar N_{n,l}}\frac{2l(l+s)}{2n+1}\frac{2\pi\Gamma(n+l+1)}{(n-l)!}\nonumber\\
&=&\iu \frac{q A_\phi}{\hbar}\frac{4 l(l+s)}{(2n+1)^2}\nonumber
\end{eqnarray}
 with the scaling given in Eq. (\ref{eq:scaling}) and $\Gamma(n+l+1)=(n+l)!$ for integers.


\section{Lowest Flat Band Density in Reciprocal Space \label{app:reciproc_dens}}
Interestingly, a simple closed form solution $\tilde{n}_0(k)$, given in Eq. (\ref{eq:lowest_charge_dens})  can be derived for the 2D radial Fourier transform by utilizing the radial symmetry of $n_0(\rho)$ that reduces the problem to a 2D Hankel transformations instead, i.e.
\begin{eqnarray}
\tilde{n}_0(k)&:=&2 \pi \int_0^\infty J_0(k r) n_0(r) r dr\nonumber\\
&=&2 \pi \int_0^\infty J_0(k r) \frac{q^2 A_\phi(1-e^{-\frac{4 qA_\phi}{\hbar} r})}{\pi\hbar  r} r dr\nonumber\\
&=&\frac{2q^2 A_\phi}{\hbar}\bigg(\frac{1}{k}-\frac{1}{\sqrt{k^2+(\frac{4qA_\phi}{\hbar})^2}}\bigg). 
\end{eqnarray}
where J denotes the Bessel function with $J_0(0)=1$ and the following Hankel relations were used for $n\mapsto\tilde{n}$ by setting $t=0$,\cite{papoulis1968systems}
\begin{eqnarray}
\frac{1}{r}&\mapsto&\frac{1}{k}\\
\frac{1}{r}J_0(t r) e^{-sr}&\mapsto&\frac{2}{\pi \sqrt{(k+t)^2+s^2}}K\bigg(\sqrt{\frac{4k t}{(k+t)^2+s^2}}\bigg).\nonumber
\end{eqnarray}
The complete elliptic integral of the first kind is labeled by $K$ with $K(0)=\frac{\pi}{2}$.

\section{Landau Levels \& Zero Currents\label{app:landau}}

Free electrons in a 2D material in the presence of a classical homogeneous magnetic field along the $z$ direction $\vec{B}_{\rm{ext}}=B\vec{e}_z$ of strength $B$ are described by the minimally-coupled Schr\"{o}dinger Hamiltonian
\begin{eqnarray}\label{Hamiltonian LLs}
\hat{H}=\frac{1}{2m_{\textrm{e}}}\left(\mathrm{i}\hbar \mathbf{\nabla}+e \mathbf{A}_{\textrm{ext}}(\vec{r})\right)^2,
\end{eqnarray}
where in the Landau gauge the external vector potential which gives rise to the magnetic field is $\mathbf{A}_{\textrm{ext}}(\mathbf{r})=-\mathbf{e}_x B y$. The Landau gauge is very convenient because it preserves translational invariance in the $x$ direction. This implies that the Hamiltonian of Eq.~(\ref{Hamiltonian LLs}) commutes with the translation operator for the $x$ direction and consequently the eigenfunctions of $\hat{H}$ in $x$ will be plane waves
\begin{eqnarray}
\phi_{k_x}(x)=e^{\textrm{i}k_x x}  \;\;\; \textrm{with}\;\;\; k_x \in \mathbb{R}.
\end{eqnarray}
Applying $\hat{H}$ on the plane waves above we have
\begin{eqnarray}\label{HLL on plane waves}
\hat{H}\phi_{k_x}=\left[-\frac{\hbar^2}{2m}\frac{\partial^2}{\partial y^2}+ \frac{m\omega^2_c}{2}\left(y+\frac{\hbar k_x}{eB}\right)^2 \right]\phi_{k_x},
\end{eqnarray}
where we introduced also the cyclotron frequency $\omega_c$
\begin{eqnarray}\label{cyclotron frequency}
\omega_c=\frac{eB}{m}.
\end{eqnarray}
In the equation~(\ref{HLL on plane waves}) the part depending on the variable $y$ remains to be treated. The part of $\hat{H}$ depending on $y$ is a shifted harmonic oscillator
\begin{eqnarray}
\hat{H}_y=-\frac{\hbar^2}{2m}\partial^2_y+ \frac{m\omega^2_c}{2}\left(y+\frac{\hbar k_x}{eB}\right)^2
\end{eqnarray}
and the eigenfunctions of the operator above are Hermite functions of the variable $y+\hbar k_x/eB$
\begin{eqnarray}
\psi_n\left(y+\frac{\hbar k_x}{eB}\right)&=&\frac{1}{\sqrt{n! 2^n}}\left(\frac{m\omega_c}{\pi \hbar}\right)^{1/4}e^{-\frac{\omega_c}{2\hbar}\left(y+\frac{\hbar k_x}{eB}\right)^2}\nonumber\\
&&\cdot H_n\left(\sqrt{\frac{m\omega_c}{\hbar}}\left(y+\frac{\hbar k_x}{eB}\right)\right)
\end{eqnarray}
with eigenvalues $\hbar\omega_c(n+1/2)$
\begin{eqnarray}
\hat{H}_y\psi_n\left(y+\frac{\hbar k_x}{eB}\right)= \hbar\omega_c\left(n+\frac{1}{2}\right)\psi_n\left(y+\frac{\hbar k_x}{eB}\right),  
\end{eqnarray}
with $n\in \mathbb{N}$. Thus, applying now $\hat{H}\phi_{k_x}$ on the shifted Hermite functions $\psi_n\left(y+\hbar k_x/eB\right)$ we obtain 
\begin{eqnarray}
\hat{H}\phi_{k_x}\psi_n=\left[\hbar\omega_c\left(n+\frac{1}{2}\right)\right]\phi_{k_x}\psi_n.
\end{eqnarray}
From the expression above we deduce that the full set of eigenfuctions for an electron in a classical homogeneous magnetic field is
\begin{eqnarray}
\Psi_{k_x,n}(\textbf{r})=\phi_{k_x}(x)\psi_n\left(y+\frac{\hbar k_x}{eB}\right),
\end{eqnarray}
with eigenenergies
\begin{eqnarray}\label{3D Landau levels}
E_{n,k_x}=\hbar\omega_c\left(n+\frac{1}{2}\right) \;\; \textrm{with}\;\; k_x, k_z\in\mathbb{R}, \; n\in\mathbb{N}.
\end{eqnarray}

Next we would like to compute the current of each Landau level. The current operator is 
\begin{equation}
    \hat{\bf{J}}=\frac{e}{m}\left(-\textrm{i}\hbar\nabla-e\textbf{A}_{\textrm{ext}}(\bf{r})\right)=\frac{e}{m}\left(-\textrm{i}\hbar\nabla+eBy\mathbf{e}_x\right)
\end{equation}
Then for the current operator on each Landau level we have 
\begin{eqnarray}
&&\frac{m}{e}\langle \Psi_{k_x,n}|\hat{\textbf{J}}|\Psi_{k_x,n}\rangle=-\textrm{i}\hbar\mathbf{e}_y \int\limits^{\infty}_{-\infty} dy \psi_{n}\left(y+\frac{\hbar k_x}{eB}\right)\partial_y \psi_{n}\left(y+\frac{\hbar k_x}{eB}\right)\nonumber \\
&&+\mathbf{e}_x\int^{\infty}_{-\infty} dy\psi_{n}\left(y+\frac{\hbar k_x}{eB}\right)\left(\hbar k_x +eBy\right)\psi_{n}\left(y+\frac{\hbar k_x}{eB}\right).
\end{eqnarray}
To compute the integrals above we introduce the coordinate $s=y+\hbar k_x/eB$ and we have
\begin{eqnarray}
\frac{m}{e}\langle \Psi_{k_x,n}|\hat{\textbf{J}}|\Psi_{k_x,n}\rangle&=&-\textrm{i}\hbar\mathbf{e}_y \int\limits^{\infty}_{-\infty} ds \psi_{n}(s)\partial_s \psi_{n}(s)\nonumber\\
&&+\mathbf{e}_xeB\int^{\infty}_{-\infty} ds\psi_{n}(s)s\psi_{n}(s).
\end{eqnarray}
Further, we use the recursion relations of the Hermite functions 
\begin{eqnarray}
    &\partial_s\psi_{n}(s)=\sqrt{\frac{n}{2}}\psi_{n-1}(s)-\sqrt{\frac{n+1}{2}}\psi_{n+1}(s)\\
     &s\psi_{n}(s)=\sqrt{\frac{n}{2}}\psi_{n-1}(s)+\sqrt{\frac{n+1}{2}}\psi_{n+1}(s)
\end{eqnarray}
and the orthogonality relations of the Hermite functions $\langle\psi_n|\psi_m\rangle=\delta_{nm}$ and we find that expectation value of the current operator for every Landau level is zero
\begin{equation}
    \langle \Psi_{k_x,n}|\hat{\textbf{J}}|\Psi_{k_x,n}\rangle=0.
\end{equation}
Finally, we would like to note that that in the presence of external constant electric field, the Landau levels get shifted by the electric field and form edge states which lead to the famous integer quantum Hall effect.~\cite{klitzing1980new, LaughlinPRB} 
\end{document}


\preprint{AIP/123-QED}

\title{SUPPORTING INFORMATION: New Class of Landau Levels and Hall Phases in a 2D Electron Gas Subject to an Inhomogeneous Magnetic Field: An Analytic Solution
}
\author{Dominik Sidler}
  \email{dsidler@mpsd.mpg.de}
 \affiliation{Max Planck Institute for the Structure and Dynamics of Matter and Center for Free-Electron Laser Science, Luruper Chaussee 149, 22761 Hamburg, Germany}
\affiliation{The Hamburg Center for Ultrafast Imaging, Luruper Chaussee 149, 22761 Hamburg, Germany}

\author{Vasil Rokaj}
  \email{vasil.rokaj@cfa.harvard.edu}
    \affiliation{Max Planck Institute for the Structure and Dynamics of Matter and Center for Free-Electron Laser Science, Luruper Chaussee 149, 22761 Hamburg, Germany}
  \affiliation{ITAMP, Harvard-Smithsonian Center for Astrophysics, Cambridge, MA 02138, USA}

\author{Michael Ruggenthaler}
  \email{michael.ruggenthaler@mpsd.mpg.de}
  \affiliation{Max Planck Institute for the Structure and Dynamics of Matter and Center for Free-Electron Laser Science, Luruper Chaussee 149, 22761 Hamburg, Germany}
  \affiliation{The Hamburg Center for Ultrafast Imaging, Luruper Chaussee 149, 22761 Hamburg, Germany}

\author{Angel Rubio}
  \email{angel.rubio@mpsd.mpg.de}
  \affiliation{Max Planck Institute for the Structure and Dynamics of Matter and Center for Free-Electron Laser Science, Luruper Chaussee 149, 22761 Hamburg, Germany}
    \affiliation{The Hamburg Center for Ultrafast Imaging, Luruper Chaussee 149, 22761 Hamburg, Germany}
  \affiliation{Center for Computational Quantum Physics, Flatiron Institute, 162 5th Avenue, New York, NY 10010, USA}
  \affiliation{Nano-Bio Spectroscopy Group, University of the Basque Country (UPV/EHU), 20018 San Sebasti\'an, Spain}

\date{\today}

\maketitle

\begin{figure*}
     \centering
     \begin{subfigure}{0.45\textwidth}
         \caption{}
        \centering
         \includegraphics[width=84.5mm]{figures/Heatmap_Hall_Conductivity_Ef_nmax_50_Eres_150_rlim_50.0cond_rx_eq_ryFalse_phi_-1.5707963267948966dnapprox_FalsecircularpertFalse_spintype_0.pdf}
        \label{fig:hallcondloc}
     \end{subfigure}%
     \begin{subfigure}{0.45\textwidth}
         \caption{}
         \centering
         \includegraphics[width=65mm]{figures/Integrated_Hall_Conductivity_density_Ef_nmax_50_Eres_150_rlim_50.0spin_0.5dnapprox_FalsecircularpertFalse_spintype_0.pdf}
         \label{fig:hallcondint}
     \end{subfigure}
        \caption{(a) Radially resolved Hall conductivity $\sigma_{xy}$ for a homogeneous electric field perturbation along the $y$-direction for $g_s=0$. To reach a locally converged solution within $r_{max}=50$ nm, a large number of single electron eigenstates with $\langle r\rangle\gg r_{max}$ is required, which indicates that the locally observed Hall conductivity switch is driven by the constant anisotropic vector potential instead of the localised inhomogeneous magnetic field. Thus, the locally resolved result is almost identical to the spin-dependent $g_s=2$ calculation, confirming that the asymptotic properties are already present at $r_{max}=50$ nm  (b) Consequently, the integrated Hall conductivity transition at $E_F=E_{A^2}(1-1/4)$ is again considered to persist in the asymptotic limit $r_{max}\rightarrow\infty$. }
\end{figure*}

\begin{figure*}
     \centering
     \begin{subfigure}[b]{0.6\textwidth}
         \centering
         \includegraphics[width=\textwidth]{figures/current_energy_radius__nmax_75_alpha_0.015_spin_0.5.pdf}
         \label{fig:expect_cur}
     \end{subfigure}
        \caption{Expected single electron currents with respect to their energy and radial position expectation values. They nicely illustrate that every electron leads has a positive total current $J^{\rm tot}\geq 0$}
\end{figure*}